\documentclass[12pt,a4paper]{article}

\usepackage{hyperref}   
\usepackage{amssymb}
\usepackage{graphicx}
\usepackage{mathtools}
\usepackage{graphicx}
\usepackage{amsmath}
\usepackage{slashed}
\usepackage{multirow}
\usepackage{amssymb,amsmath,amsfonts}
\usepackage[sort&compress,numbers]{natbib}

\graphicspath{{figures/}}

\usepackage{hyperref}
\usepackage{amssymb}
\usepackage{graphicx}
\usepackage{xcolor,amsmath,amsfonts,graphicx,amssymb,subfigure}
\usepackage{graphicx}
\usepackage{authblk}

\graphicspath{{figures/}}

\graphicspath{{./figures/}}

\graphicspath{{./figures/}}

\begin{document}

\title{Gliding down the  QCD transition line,\\ from $N_f=2$ till the onset of conformality}
\author[1,2]{Andrey Yu. Kotov\thanks
{\tt a.kotov@fz-juelich.de}} 
\author[3]{Maria Paola Lombardo\thanks{\tt lombardo@fi.infn.it}}
\author[4]{~~~~~Anton Trunin\thanks{\tt amtrnn@gmail.com}}

\affil[1]{Juelich Supercomputing Centre, Forschungszentrum Juelich, D-52428 Juelich, Germany}
\affil[2]{Bogoliubov Laboratory of Theoretical Physics, Joint Institute for Nuclear Research, Dubna, 141980 Russia}
\affil[3]{INFN, Sezione di Firenze, 50019 Sesto Fiorentino (FI), Italy}
\affil[4]{Samara National Research University, Samara, 443086 Russia}
\maketitle

\begin{abstract} {We review the hot QCD transition with varying number of flavors, from two till the onset of the conformal window. We discuss  the universality class
for $N_f=2$, along the critical line for two massless light flavors, and a third flavor whose mass serves as an interpolator
between $N_f = 2$ and $N_f=3$. We identify a possible scaling window for the 3D $O(4)$ universality class transition,
and its crossover to a mean field behaviour.  We follow the transition from $N_f=3$ to   larger $N_f$, when it remains  of first order, with an increasing coupling strength; we summarize its known properties,  including possible cosmological applications as a model for a strong electroweak transition. The first order transition, and its accompanying second order endpoint, finally morphs into the essential singularity at the onset of the conformal window, following the singular behaviour predicted by the Functional Renormalization Group.}
\end{abstract}

\section{Phases of QCD and critical behaviour}

Strong interactions have different phases in the space of the number of flavors~$N_f$, quark mass, temperature~\cite{Pisarski:1983ms,Miransky:1996pd}. At low temperatures and low number of flavors their chiral symmetry is spontaneously broken. The hot symmetric phase is known as quark gluon plasma; in the chiral limit the phase transitions may be of a second order for $N_f=2$, probably in the universality class of the three dimensional $O(4)$ ferromagnet.  The addition of a third flavor to the $N_f=2$ theory produces the so-called $N_f=2+1$ theory, which interpolates
between $N_f=2$ and $N_f=3$~\cite{Rajagopal:1992qz}. The strength of the transition increases with $N_f$~\cite{Shuryak:2013bxa}, and it is unclear when it turns into a first order transition~\cite{Philipsen:2019rjq,Cuteri:2018wci,Cuteri:2017gci}. At zero temperature the symmetric phase is conformal: it is separated from the broken phase by a conformal phase transition\cite{Miransky:1996pd, MIRANSKY_2010} - similar to a Berezinskii--Kosterlitz--Thouless (BKT) transition:the scaling of the order parameter reveals an essential singularity.  It is not clear - to our knowledge - how the line of first order phase transitions expected at large $N_f$  would turn into  a conformal transition, and indeed other scenarios are possible,  including a power-law scaling \cite{Braun:2010qs} and even a first order transition \cite{Antipin:2012sm,Sannino:2012wy}.

\begin{figure}[t]
    \vskip -4 cm
\hskip -1 cm     \includegraphics[width=15cm]{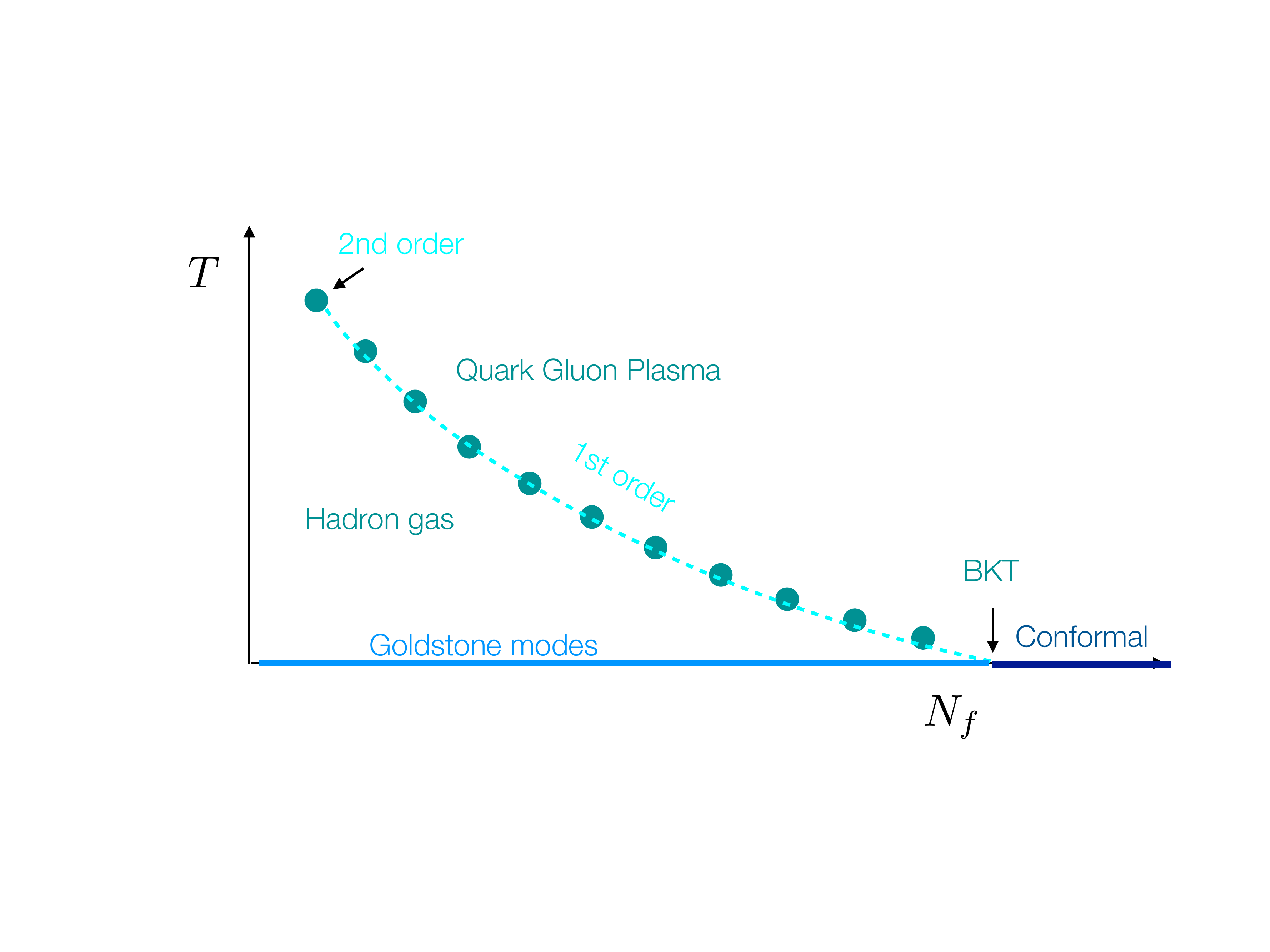}
    \vskip -2 cm
   \caption{Sketchy view of the phases of strong interactions in the space spanned by
    $N_f$ massless flavors, and temperature $T$.}
    \label{fig:tnf}
\end{figure}

The critical line of QCD (Figure~\ref{fig:tnf}) separates the hadronic phase from 
a hot phase where chiral symmetry is restored - for physical values of the quark masses, this is the phase explored in heavy ion collisions, much explored also on the lattice \cite{Ratti:2018ksb,Ding:2020rtq}. At zero temperature,
in the broken phase, we have the Goldstone singularity. Above a critical
number of flavors the theory is conformal, with anomalous dimension~\cite{Miransky:1996pd}.
The global symmetry of QCD: 
$U(n)_L\!\times\! U(n)_R \cong SU(n)\!\times\! SU(n)\!\times\! U(1)_V\!\times\! U(1)_A$ 
valid at classical level is broken by topological fluctuations, for which the
$\eta'$ mass gives an experimental evidence. 
The remaining symmetry is then 
$U(n)_L\!\times\!U(n)_R / U(1) \cong SU(n)\!\times\!SU(n)\!\times\! U(1)_V$. 
This prompted the question \cite{Shuryak:1993ee}: {\em Which chiral symmetry is restored at high temperature?}
$U(1)_A$ will always be broken, but the amount of breaking may well be sensitive to the temperature, leading to an approximate restoration, and a natural question arises on the interrelation of the $SU(N) \times SU(N)$
symmetry with the $U(1)_A$ symmetry. 
Since the chiral condensate breaks the $U(1)_A$ symmetry, the only possibilities are a near-coincidence of the two transitions, or an axial breaking persisting beyond chiral restoration.


The axial symmetry is discriminating: 
if its breaking is not much sensitive
to the chiral restoration, the breaking pattern
for $N_f=2$ is indeed 
$SU(2)_L\!\times\!SU(2)_R \to SU(2)_V$  or
$O(4) \to O(3)$~\cite{Pisarski:1983ms}. Due to the associate diverging correlation length, the theory is effectively three dimensional, leading to the well known 3D $O(4)$  universality class.  If instead axial symmetry is correlated with chiral symmetry, the relevant breaking pattern is
$U(2)_L\!\times\!U(2)_R \to U(2)_V$, hinting either at a first 
or
even at a second order transition with different exponents~\cite{Pelissetto:2013hqa}. 

Beyond two flavors, the issue of the anomaly becomes more subtle: the definition of a proper order parameter for axial symmetry is entangled with different susceptibilities associated with different flavors \cite{Nicola:2020wxy}. Some studies indicate restoration above $T_c$ ~\cite{Ding:2020xlj,Kaczmarek:2021ser,Kaczmarek:2020sif, Aoki:2021qws,Aoki:2020noz,Mazur:2018pjw,Buchoff:2013nra,Suzuki:2020rla,Kanazawa:2015xna,Aoki:2012yj,Tomiya:2016jwr,Brandt:2019ksy,Brandt:2016daq,Cossu:2013uua,Chiu:2013wwa},
others find hints of a near-coincidence of the two transitions \cite{Philipsen:2019rjq,Brandt:2016daq}. Our recent study~\cite{Kotov:2021rah}, which will be reviewed in detail
in Section~\ref{sec:nf2p1}, attempts at quantifying the limit of the scaling window and finds compatibility with 3D $O(4)$, thus implicitly suggesting a separation between the two transitions. However, 
we have also observed a correlation between the $\eta'$ meson mass and the chiral condensate around the transition, which may also be compatible with their coincidence~\cite{Kotov:2020hzm,Kotov:2019dby}. Figure \ref{fig:2ndtor}
and Figure \ref{fig:1stor} illustrate two possible scenarios for the critical behaviour and scaling window between $N_f=2$ and $N_f=3$. We will discuss them in detail in Sections 3 and 4. 

For $N_f = 3,4$ the standard lore is a first order transition, even if some contrasting
evidence has been reported \cite{Philipsen:2019rjq}. The strength of the transition increases with $N_f$
\cite{Lombardo:2014mda,Miura:2012zqa,Miura:2011mc,Shuryak:2013bxa},  and this has been used as
a possible paradigm for the generation of gravitational waves at a strong electroweak transition in models with composite Higgs 
\cite{Cacciapaglia:2020kgq}. 

All the phenomena above are intrinsically non-perturbative, and the lattice approach has been extensively used to address them. They are often discussed from different viewpoints, having in mind different applications. Here, we would like to present a general overview, attempting at a synthesis.
The remaining of this  report is organised as follows: in the next Section we review
the theoretical knowledge about the critical line.  The following two Sections contain results for $N_f=2$ and $N_f=2+1$. In these Sections we rely mostly on our 
work, and, for the latter case, we include some unpublished analysis. In addition, we use this case to illustrate some recent proposal for the study of the critical behaviour. 
Section~\ref{sec:nf3-4} reviews the effort towards the identification of the critical endpoint of a first order transition for $N_f = 3,4$. Section~\ref{sec:largenf} is devoted to large $N_f$ and to  the approach to the conformal window. We conclude with a brief summing up. 

\section{Universal approach to phase transitions} 
\label{sec:sec2}

We summarize here a few general aspects of the different critical behaviours encountered along the critical line, while the numerical
evidence for the different possibilities is discussed in the following Sections. 

To make this discussion self-contained, let us summarize a few facts about phase transitions and critical behaviour, see e.g. Ref.~\cite{Pelissetto:2000ek} for a complete discussion. We consider  a system undergoing a phase transition between phases characterised by different symmetries, under the action of an external parameter (temperature, for instance). Early descriptions of such systems were made in the framework  of the Landau mean-field theory, which is based on a local, space homogeneous order parameter $M$. The free energy $F$ is analytic in $M$
and in the temperature $T$, and it is truncated   to fourth order in $M$: $F(M,T) = F(0,T) + V a(T)M^2 + V b(T)M^4$, with $a(T) = a_0 \tau$ and $b= b_0$, and $a_0,b_0$ are positive. $\tau$ is the reduced temperature $\tau = (T - T_c)/T$. 
Under these assumptions,
the miniminization of the free energy gives the well-known power-law behaviour for the order parameter with  $M(T) = M_0 \tau ^\beta $, $\beta = 1/2$. The Landau theory is readily generalised to include an external field linearly coupled to the order parameter,
$F(M,T) = F(0,T) + V a(T)M^2 + V b(T)M^4 -V M h$.
The power-law singularity at $h=0, T=T_c$ is washed out, while a singular behaviour at $T_c$ is manifest in the scaling of the order parameter
$M \propto h^{\delta}$, $\delta = 1/3$. 
Experiments, however, show that the mean field exponents are not accurate: to address this, a phenomenological scaling theory has been developed, which still produces a power-law 
behaviour for the order parameter, but with different exponents. A pivotal assumption, 
theoretically motivated within a Renormalization Group approach, is that the behaviour of the system is completely controlled by a diverging correlation length at the critical point.
The essence of the behaviour is captured by the universal Equation of State (EoS), which is characteristic of a given combination of symmetry breaking pattern and dimensionality:
\begin{equation}
    M/h^{1/\delta} = f(t/h^{1/\beta\delta}).
\end{equation}
In the QCD EoS  we will identify  $M \equiv \bar \psi \psi$,   $h \equiv m_q$,  
$t \equiv  T - T_c$,  $m_q$ is the quark mass,
and $T_c$ is the critical temperature in the chiral limit: the bare quark mass and the chiral condensate 
play the role of the external breaking field and of the spontaneous magnetization. 
Note that there are two arbitrary normalizations for $M$
and for $T$. 
A detailed discussion together with explicit calculations in spin models may be found e.g. in Ref.~\cite{Engels:1999wf}. $f$ is a regular function: by expanding it to first order, and setting $\beta = 0.5, \delta = 3$ one recovers the Landau mean field behaviour. 
The question now is, what triggers the crossover from  mean field  to the critical behaviour? A short answer is to follow the Ginzburg criterium~\cite{Ginzburg}: the correlation length increases towards the critical point, and at some point the fluctuations take over, the details of the microscopic behaviour do not matter, and the system shows the appropriate universal behaviour. Interestingly, the same reasoning applies to weakly first order transitions~\cite{Fernandez:1992ns}. 
In short summary, when approaching a critical region, one may observe first a mean-field behaviour, then, when the Ginzburg criterium is satisfied, the true critical behaviour will appear.
The crossover between the interaction-dominated region, which follows  mean-field predictions, to the true critical regime, dominated by the diverging correlation length,   has been extensively 
studied in condensed matter systems~\cite{Hohenberg_2015, Pelissetto:2013hqa}. In the following, we will search for it in the QCD transition where it is much less explored.

{\em Let us consider first the case of a continuous, second order transition.} The discussion is general, we will use, however, as concrete examples the mean field and the three dimensional $O(4)$ universality class.

To describe the critical behaviour it is convenient
to use an alternative, equivalent  form of the EoS for the order parameter:
\begin{equation}
M = h^{1/\delta}f_G(t/h^{1/{\beta \delta}}).
\end{equation}

The high $x$ and low $x$ expansions 
\begin{eqnarray}
f_G(x) &=& x^{-\gamma} \sum_{n=0}^\infty d_n x^{-2n \Delta}, x \to + \infty \\
&=&(-x)^\beta \sum_{n=0}^{\infty} c_n (-x)^{-n \Delta/2}, x \to - \infty
\end{eqnarray}
with $x \equiv t/h^{1/{\beta \delta}} $,
 $\Delta \equiv \beta \delta$, $\gamma = \beta (\delta -1)$
are known~\cite{Engels:2011km}, and the coefficients have been computed in spin models for the $O(4)$ continuous universality class~\cite{Engels:2011km}. 
Ref.~\cite{Engels:2011km} found a good  interpolating form  around $x = 0$:
\begin{equation}f_G' (x)=b_1 +2b_2x+3b_3x^2 +4b_4x^3 +5b_5x^4 +6b_6x^5,
\end{equation}
whose coefficients are tabulated in the paper~\cite{Engels:2011km}.

To identify the critical scaling, and the critical temperature in the chiral limit, at finite temperatures there are basically three (interrelated) strategies:
\begin{itemize}
    \item direct comparison  with the Equation of State
    \item the study of the dependence of the pseudo-critical temperatures on the breaking field, also known as scaling of  pseudo-critical temperatures
    \item definition of RG invariant quantities, which do not depend on the breaking field at the critical point.
\end{itemize}

The second one is probably the most popular: in practice,  
one  relies on pseudo-critical temperatures associated with features of the 
order parameter, or related observables. For instance, 
considering the expression for the  susceptibilities  
\begin{eqnarray}
\chi_L &=& \frac{\partial \bar \psi \psi}{\partial m} \nonumber, \\
\chi_\Delta &=&  \frac{\partial \bar \psi \psi}{\partial T}
\end{eqnarray}
derived from the EoS, one finds that for the $O(4)$
universality class they
 peak  at 
$t/h^{1/\beta \delta} = 1.35(3)$  
and $t/h^{1/\beta \delta} = 0.74(4)$, respectively. 
The corresponding pseudo-critical temperatures
\begin{equation}
T_c^s(m_\pi) = T_c(0) + k_s  m_\pi^{2/\beta \delta}
\label{eq:tc}
\end{equation}
(where $s$ labels the different observables)
should scale with the pion mass $m_\pi$ with the same exponent ${2/\beta \delta}$,
 but with different $k_s's$, whose ratio is a prediction of universality.
 The  longitudinal  and transverse susceptibility $\chi_L$ and $\chi_T$, where $\chi_T \equiv \langle \bar \psi \psi\rangle/m$,
  may be used to implement the third approach, based on RG invariant quantities \cite{Kocic:1992is,Karsch:1994hm,Ding:2019prx}.
 
\begin{figure}[t]
\centering
    \includegraphics[width=12cm]{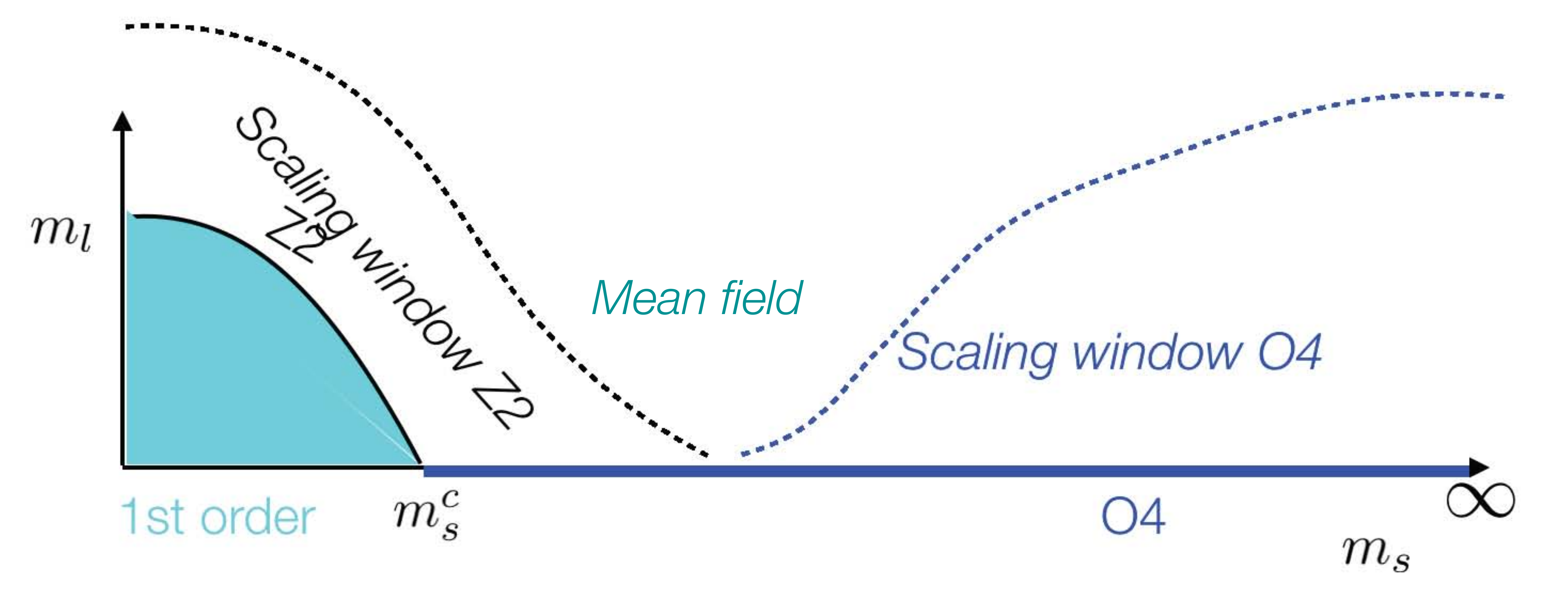}
    \caption{Zooming in the region between $N_f=2$ and $N_f=3$: assuming a 3D $O(4)$
    scenario, with  hypothesized scaling windows in the $m_l, m_s$ plane (upper diagram).The dotted lines are a possible sketchy behaviour of the crossover between the mean field region and the critical region.} 
    \label{fig:2ndtor}
\end{figure}

\begin{figure}[t]
    \centering
    \vskip -3 cm
    \includegraphics[width=15cm]{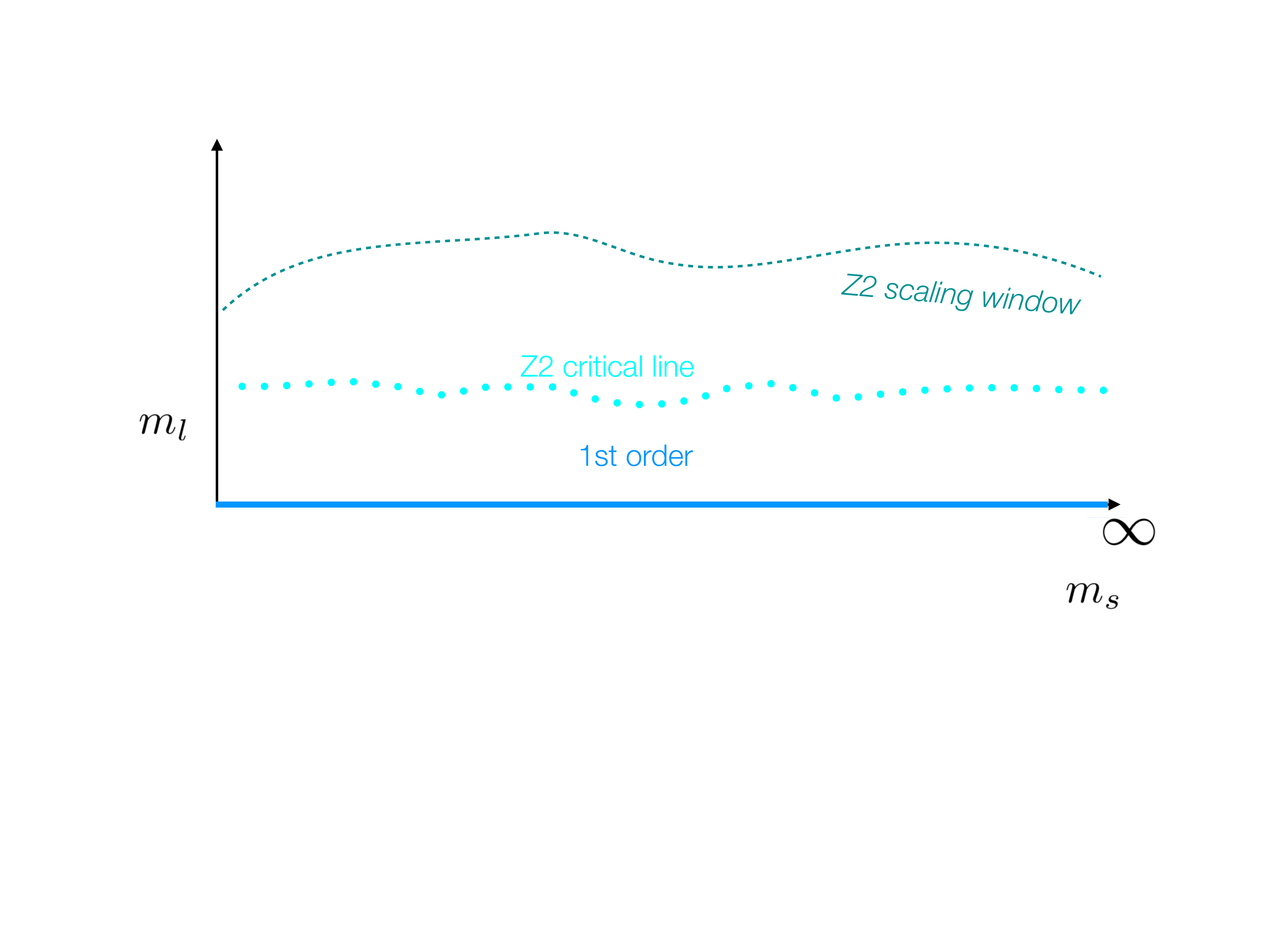}
    \vskip -3 cm
    \caption{As Figure~\ref{fig:2ndtor}, but for a first order transition extending from $N_f=2 $ to $N_f=3$. There are no theoretical predictions for the  shape of the critical $Z_2$ line and the scaling window, the lines are merely  indicative. Above the upper dotted line the behaviour should be compatible with mean field.} 
    \label{fig:1stor}
\end{figure}
 
All these approaches  are prone to suffer
from the contamination of regular terms, especially when the pseudo-critical temperature $T_c^s$ associated with the particular observable $s$ under consideration  has a strong dependence on the breaking field, i.e. on the pion mass (see also Refs.~\cite{Ding:2020xlj,Kotov:2021rah}).
These considerations suggest an alternative order
 parameter~\cite{Kotov:2021rah}, see also~\cite{Kogut:1998rh,Unger:2010wcq}, free from linear contributions:
\begin{equation}
\langle \bar\psi\psi \rangle_3  \equiv m(\chi_T - \chi_L) \equiv \langle \bar \psi \psi \rangle - m \chi_L \equiv \langle \bar \psi \psi \rangle - m  \frac{\partial \langle \bar\psi\psi \rangle}{\partial m}.
\end{equation}
We dubbed this order parameter $\langle \bar\psi\psi \rangle_3$
to highlight the fact that the leading $m$ correction in its Taylor expansion, when defined, is $m^3$. Longitudinal and transverse susceptibility become degenerate at the transition in the chiral limit, hence
their difference is an order parameter.

The $m$ factor has been included to avoid divergencies in the chiral limit in the broken phase. 
The associated Equation of State reads: 
\begin{equation}
\frac{\langle \bar\psi\psi \rangle_3}{m^{1/\delta}} = f_G(x)( 1 - 1/\delta) + \frac{x}{\beta \delta} f_G(x)'.
\end{equation}
Interestingly, the high temperature
leading term is  $\langle \bar\psi\psi \rangle_3 \propto t^{-\gamma - 2 \beta \delta}$ rather than $\langle \bar\psi\psi \rangle \propto  t^{-\gamma}$: the decay is rather fast, not surprisingly given that this observable is closer to the chiral condensate in the chiral limit.

\begin{figure}[t]
\includegraphics[width=13truecm]{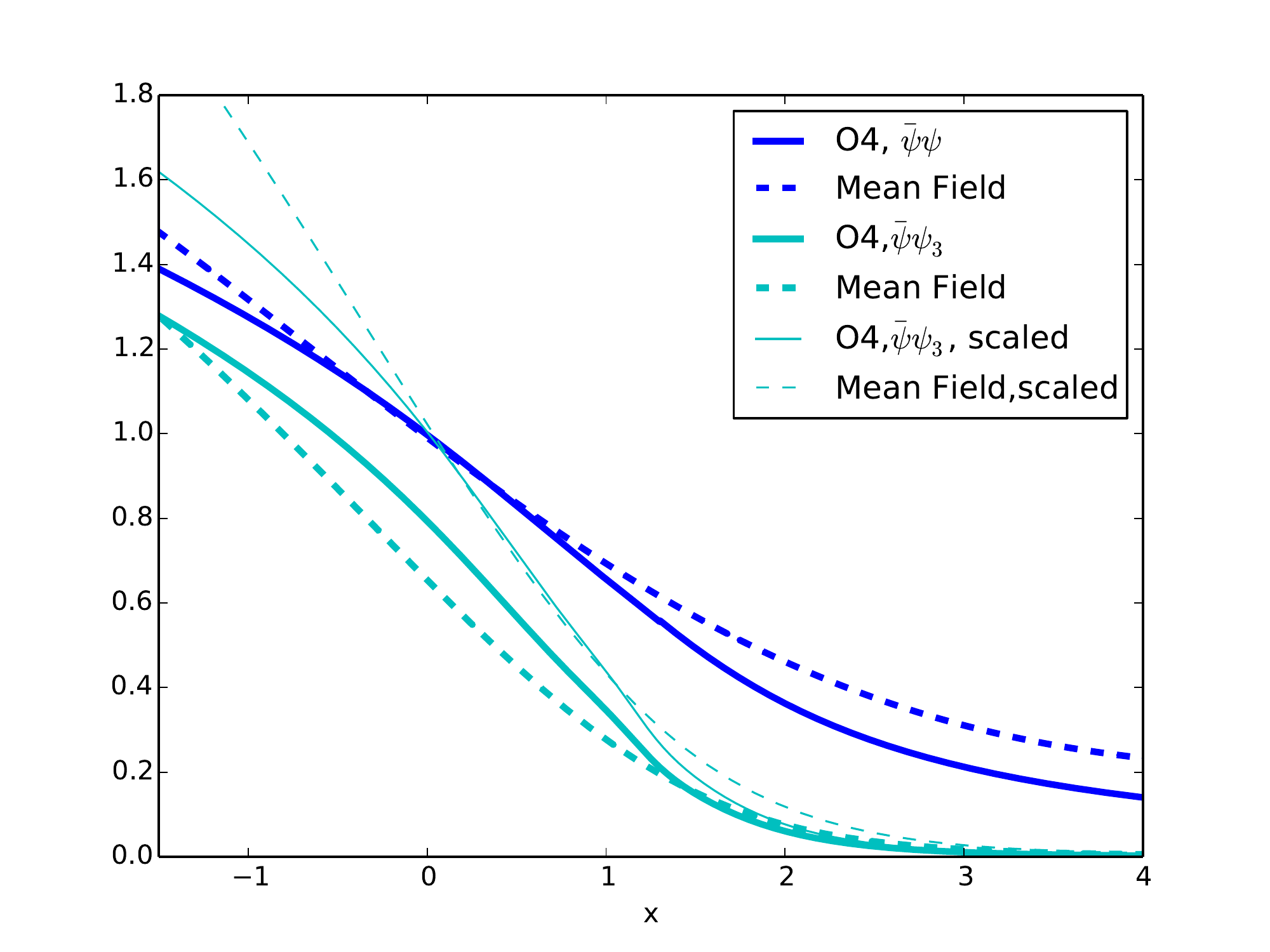}
\caption{The Equation of State for the chiral condensate $\langle \bar\psi\psi \rangle$ and the new order parameter $\langle \bar\psi\psi\rangle_3$ in the critical region for the $O(4)$ three dimensional universality class, and for mean field. For a more direct comparison we also plot the results 
suitably rescaled as thin lines (from Ref.~\cite{Kotov:2021rah}).}
\label{fig:O4_pbp}
\end{figure}

In Figure~\ref{fig:O4_pbp} 
we compare the EoS for $\langle \bar\psi\psi \rangle _3$ with the one for 
$\langle \bar \psi \psi \rangle$
for the 3D $O(4)$ Universality class, and for mean field.   Note
the sharper decrease of $\langle \bar \psi\psi \rangle_3$, consistent with
it being closer to the critical behaviour. 
Away from criticality  dimensional reduction is
less and less justified, and the system remains four dimensional and
possibly closer to mean field.  For instance, mean field scaling has been reported in large-$N$ Gross-Neveu~\cite{Kocic:1995nf}, where the scaling window shrinks to zero, and also in weak first order transitions \cite{Fernandez:1992ns}. The extent of the scaling window is a non-universal feature - a recent analysis for spin models is in Ref.~\cite{Caselle:2020tjz}. It is then very natural to compare
the 3D $O(4)$ Equation of State with the prediction of mean field:
mean field  is indeed very close to 3D $O(4)$ (see again Figure~\ref{fig:ealnf}), so the transition from the scaling window to a regime with small fluctuations could be very smooth. 

From the Equation of State data we can estimate 
the inflection point, which will drive the behaviour
of the pseudo-critical temperature associated with $\langle\bar\psi\psi\rangle_3$, $x_\text{infl} = 0.55(1)$
where the error has been estimated from the dispersion of different fits interpolating the high and low temperature branches. Table~\ref{tab:ks} summarizes the finding for
the $k_s's$ for the different chiral observables. 

\begin{table}
\centering
\begin{tabular}{c|c|c|c}
\hline
Observable  & $\chi$ & $\langle \bar\psi\psi \rangle$ & $\langle \bar \psi\psi \rangle_3$ \\
\hline
$k_s$         & 1.35(3) & 0.74(4) & 0.55(1) \\
\hline
\end{tabular}
\caption{$k_s$ for three chiral observables, from the 3D $O(4)$ Equation of State, see Eq.~\eqref{eq:tc}.}
\label{tab:ks}
\end{table}

As we will discuss in Section~\ref{sec:nf2p1}, as of today, $N_f =2 $  is serious candidate for a second order behaviour.

{\em We move from second to first order transition} by increasing
$N_f$. One way to interpolate continuously between
different $N_f$'s is by tuning the mass of the 'extra' flavor. The original discussion is Ref.~\cite{Rajagopal:1992qz}, and refers to the horizontal axis of Figure~\ref{fig:2ndtor}: there is a first order transition for
$N_f=3$, terminating at a critical point in the $Z_2$ universality class
at  $m_s = m_\text{crit}^s$. For $m_s \gg m_\text{crit}^s$,
 $m_s$ merely renormalizes the coefficients of the effective action, resulting in a shift
of the critical temperature, without changing the critical behaviour~\cite{Rajagopal:1992qz}.
In this case one conventionally assumes that there is a line of second
order transition $\infty > m_s > m_\text{crit}^s$, $T_c=T_c(m_s)$. 
The question is, how the scaling window for $N_f=2$ morphs into the scaling window 
around $m_\text{crit}^s$.  Figure~\ref{fig:2ndtor} presents a  simplistic scenario: the scaling windows in $m_l$ on either sides shrink till they almost disappear in the middle. So the two scaling windows basically do not communicate. A more compelling answer would require an analysis of
the pseudo-critical behaviour around $m_\text{crit}^s$~\cite{Pisarski}. 
Interestingly, in Ref.~\cite{Nicola:2020wxy} the standard 
subtracted condensate 
\begin{equation}
\chi_S -\chi_K = \frac {2 m_s}{m_s^2 - m_l^2}\bigl[\langle \bar q q \rangle_l(T) - 2 \frac{m_l}{m_s} \langle\bar s  s\rangle (T)\bigr]
\end{equation}
has been advocated as a diagnostic
tool for the behaviour with a finite $m_s$.
Figure~\ref{fig:1stor} shows the alternative first order scenario,
which is also a generic prototype for larger $N_f$.

The {\em first order}  region for larger $N_f$ is 'uneventful' from the perspective of the critical behaviour. Its important feature is the
endpoint: when the breaking field becomes stronger, the transition weakens, and finally it becomes a continuous one. The weakening of the first order transition has been studied in detail in q-state Potts models \cite{Fernandez:1992ns}, where the strength of the transition has been linked
to the position of the spinodal point - the apparent divergence point
of the correlation length.  At the endpoint of the first order transition
the strength becomes zero, and the spinodal points collapse on the critical point. The axes are no longer the usual ones, and are defined
by the directions of the first order line. A clean observation of the endpoint is essential to complete the analysis of a first order behaviour. 

When $N_f$ increases, the coupling at the transition is known to become stronger \cite{Miura:2011mc,Miura:2012zqa}. The zero temperature theory has scale separation, and may be used to model a composite Higgs \cite{Cacciapaglia:2020kgq}. The high temperature first order transition  may offer 
a model of a strong electroweak transition \cite{Miura:2018dsy}, a very attractive possibility for gravitational wave generation.

{\em The  zero temperature quantum phase transition is expected to be conformal}~\cite{Miransky:1996pd}, although other possibilities cannot be excluded, including a first order transition \cite{Antipin:2012sm,Sannino:2012wy}, and a power-law scaling \cite{Braun:2010qs}. It occurs for a non-integer number of flavors, and observing it by extrapolation
needs a control on the scaling setting procedure for different theories.
The behaviour with a finite mass is less established in this case. It is studied in Ref. \cite{PhysRevLett.108.255703}, but to our knowledge 
this general scaling has not been directly applied to the case at hand. 
The universal behaviour of a conformal transition with a breaking field  remains an open problem.

\section{\texorpdfstring{$N_f=2$}{nf2}}
\label{sec:sec3}

A much discussed scenario for $N_f=2$ is a second order transition, see Figure~\ref{fig:2ndtor}.  The search for universality  is mostly done via the scaling of the pseudo-critical temperature according to Eq.~\eqref{eq:tc}. The scaling works, within the large errors: basically, the data are consistent with a linear scaling of the pseudo-critical temperature with 
the pion mass, to be compared with the predicted power law scaling
with $ T_c(m_\pi) \propto m_\pi^{2/\delta} \simeq m_\pi^{1.08}$ for the 3D $O(4)$ universality class. The $U(2) \times U(2) \to U(2)$ pattern   predicts a very similar scaling, $m_\pi^{2/\delta} \simeq m_\pi^{1.16}$,
leading to an indistinguishable behaviour within the current errors. 

The possibility of a first order transition is also explicitly considered for two flavors. In such scenario, depicted in Figure~\ref{fig:1stor},   the first transition region stretches all the way till there  $N_f=3$, bordered by  a line of $Z_2$ endpoints~\cite{Philipsen:2019rjq}.

The $Z_2$ endpoint has been extensively searched for in
QCD with three flavors (see next Section), and it has proven to be elusive and very sensitive to lattice details. As a part of these uncertainties, 
there is no clear indication of mixing at the critical point,
so in practical analysis the mixing is ignored. The search for
a first order scenario then relies on direct searches,
so far unsuccessful, at small masses, as well as on the scaling of the pseudo-critical temperature:
\begin{equation}
T_c^s(m_\pi) = T_c + k_s A (m_\pi^2 - m_c^2)^{1/{\beta \delta}}
\end{equation}
with $1/{\beta \delta} = 0.64$ for the $Z_2$ universality class
\cite{Vicari:2008jw}.

The outcome of these analysis  \cite{Burger:2011zc} is that there is no evidence for $m_c$.  A recent study \cite{Brandt:2016daq} confirms these findings, after performing a careful comparison of the different breaking patterns. Summing up, 
it is impossible to discriminate among different universality classes on the basis of the scaling of $T_c (m_\pi)$ alone. On the positive side, the critical
temperature in the chiral limit is robust  against different choices:
$T_c (0) (O(4)) = 163(27)$ MeV and $T_C (0) (U(2)\times U(2) ) = 167(25)$ MeV, which compares well with the twisted mass results
$T_c = 152(26)$ \cite{Burger:2011zc}.

We mark this result in the
$m_\pi, m_s, T$ space in Figure~\ref{fig:nf2nf3sum}, and in the
$N_f,T$ plane in Figure~\ref{fig:tcnf}, which we will discuss more later. 

On the analytic side, interesting studies in four dimensions \cite{Braun:2010qs} have suggested scaling behaviour only for pion
masses below 1 MeV. There is, however, an apparent scaling for much larger masses, and
it would be interesting to see whether the apparent scaling for larger masses is compatible with a mean field analysis.

Important complementary information comes from the analysis of 
screening masses \cite{Shuryak:1993ee}: some studies  find the axial breaking much reduced  at the chiral transition.
A detailed discussion is found in Ref.~\cite{Brandt:2019ksy},
but the issue remains open as different observables appear to give different information.

\section{\texorpdfstring{$N_f=2+1$}{nf2p1}, and the physical point} 

\label{sec:nf2p1}

This is a much studied theory, as it includes the physical case of a strange mass (see
Figure~\ref{fig:2ndtor}) with hope that the light quarks will still be within, or not too far from, the scaling window. We note that the results in the chiral limit may have a phenomenological relevance, 
according to low energy effective theory computations: the two massless
flavor chiral transition temperature is an upper bound for the temperature of the critical
endpoint~\cite{Ding:2019prx}. Clearly only a full ab-initio computation may confirm or disprove this, and, in turn, 
such observation would be a validation of these models.

This Section is mostly  based on our recent work~\cite{Kotov:2021rah}, where we have made use of the ad-hoc order parameter introduced in Section~\ref{sec:sec2}. The results are obtained with a dynamical charm. However, 
around the critical temperature a dynamical charm is completely decoupled, hence we are effectively discussing the $N_f = 2 +1$ theory, with a physical strange mass.
 We have simulated four different pion masses,
from the physical value till 470 MeV. Our simulations are performed in the  fixed scale approach, where we keep the bare lattice parameters fixed and vary temperature by varying the number of lattice spacings in the temporal direction, to cover a temperature span ranging from 120 MeV till 800 MeV, approximatively. 
Our ensembles as well as more details can be found in
Refs.~\cite{Kotov:2021rah,Kotov:2020hzm,Kotov:2019dby}.

Before turning to our results, let us
briefly summarize the current status. 
By use of a subtracted condensate and related susceptibilities, as well as finite volume scaling,  Refs.~\cite{Ding:2020xlj,Ding:2020rtq} find a satisfatory $O(4)$ scaling up to nearly physical pion mass, with $T_c = 132^{+3}_{-6}$ MeV. 
A recent FRG study~\cite{Braun:2020ada} confirms these findings,  but with a slightly larger $T_c = 142$ MeV in the chiral limit.

For the discussion of the universality class and the chiral limit
we consider the chiral condensate, the connected and the full susceptibility.  
These observables suffer from an additive renormalization, which, in our fixed scale approach,   does not  affect the estimate of the pseudocritical point. 
However, it hampers the direct comparison with the Equation of State,
and blurs the behaviour of the pseudo-critical temperatures, which receive
mass corrections. By contrast, the observable $ \langle\bar\psi\psi\rangle_3$: 
\begin{equation}
    \langle\bar\psi\psi\rangle_3 = \langle\bar\psi\psi\rangle - m_l \chi_L
\end{equation}
 is free from linear additive renormalization as well as from linear correction to scaling.


We use various functional forms 
to parameterize our observables in various intervals, and to identify
the associated {\em pseudo-critical temperature}. 
We then use the difference among results from different intervals/fitting forms to estimate the systematic error.
In some cases, in particular for the full susceptibility,  no explicit parameterization fared well through the data. In this case, we have also used  cubic splines as smooth interpolators,  estimating statistical uncertainty by adding random Gaussian noise to each point, weighted by statistical uncertainty of our data points.  The details 
can be seen in our recent publication~\cite{Kotov:2021rah}.

\begin{figure}
    \centering
    \vskip -1cm
    \includegraphics[width=15cm]{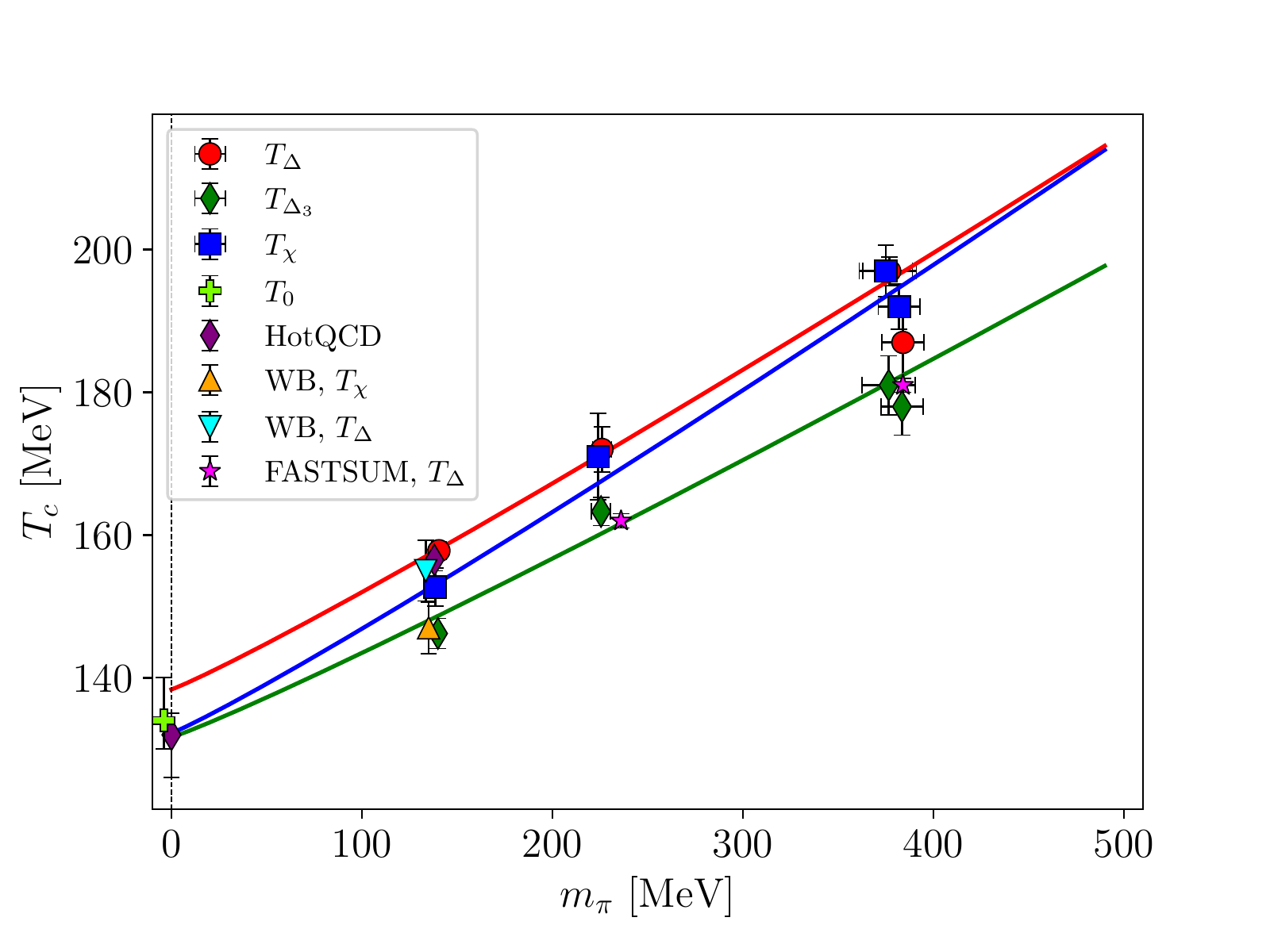}
    \caption{Pseudo-critical temperatures with their chiral extrapolations: comparison with the results from the HotQCD Collaboration~\cite{Bazavov:2018mes}, FASTSUM Collaboration~\cite{Aarts:2019hrg,Aarts:2020vyb}, Wuppertal-Budapest Collaboration~\cite{Borsanyi_2020}. The purple diamond at $m_{\pi}=0$ marks the
    critical temperature~\cite{Ding:2019prx}, which compares well with our result $T_0 = 134^{+6}_{-4}$~MeV (light-green cross, slightly shifted for better readability). From Ref. \cite{Kotov:2021rah}.}
    \label{fig:crittemp}
\end{figure}
In Table~\ref{tab:crit_temp}, reproduced from Ref.~\cite{Kotov:2021rah}, we summarize our results for the pseudo-critical temperatures extracted from different chiral observables.

\begin{table}
\begin{center}
\begin{tabular}{|c|c|c|c|}
\hline
$m_\pi$ [MeV] & $T_\Delta$ & $T_{\Delta_3}$ & $T_{\chi}$ \\
\hline
139& 157.8(7)(10)  & 146.2(21)(1) & 152.7(13)(23) \\
225& 172(3)(1) & 163.3(18)(8) & 171(6)(1) \\
383& 187(5)(1) & 178(4)(0) & 192(3)(1) \\
376& 197(2)(0) & 181(1)(4) & 197(2)(3) \\
\hline
\end{tabular}
\caption{Pseudo-critical temperature extracted from
the chiral observables, from Ref.~\cite{Kotov:2021rah}.}
\label{tab:crit_temp}
\end{center}
\end{table}

The fits for the pseudo-critical temperatures proceed exactly as for
the $N_f=2$ case, so we do not repeat the discussion here,
and simply show the summary plots, from Ref.~\cite{Kotov:2021rah}, in Figure~\ref{fig:crittemp}.
Mutatis mutandis, it remains true that the results in the chiral limit
do not depend on the universality class.

An interesting added feature is the possibility to check the ratio of the $k_s's$: the scaling  is not quantitatively accurate, but 
to some extent consistent with 3D $O(4)$.

\begin{figure}[t]
    \centering
    \includegraphics[width=10cm]{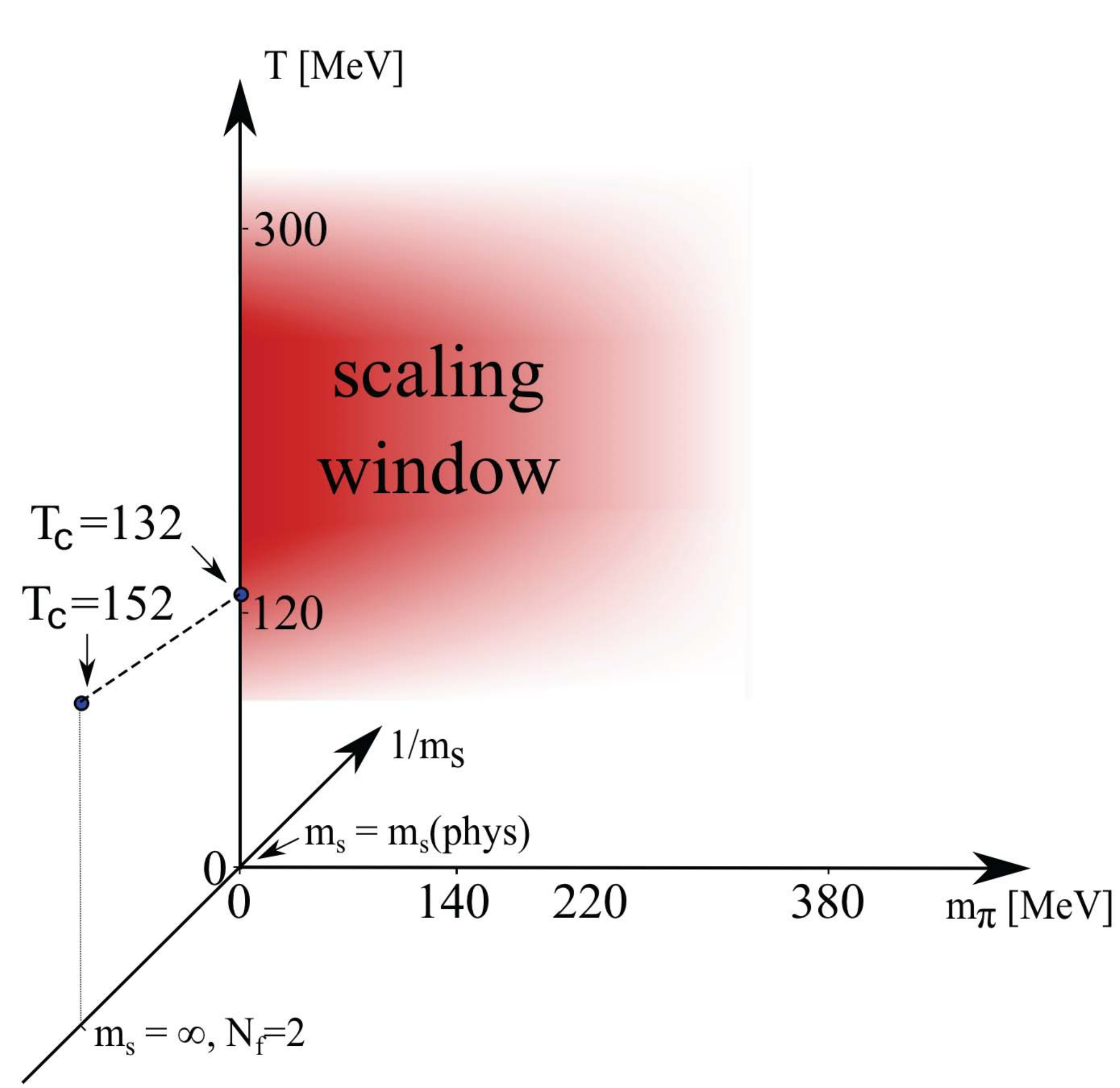}
    \caption{The space spanned by the pion mass, the strange mass, and the temperature, for strange masses ranging from infinite till the physical value. The scaling window identified for $N_f=2+1$ is marked in shades of red.}
    \label{fig:nf2nf3sum}
\end{figure}

\begin{figure}[t]
    \centering
    \vskip -3cm
    \includegraphics[width =12cm]{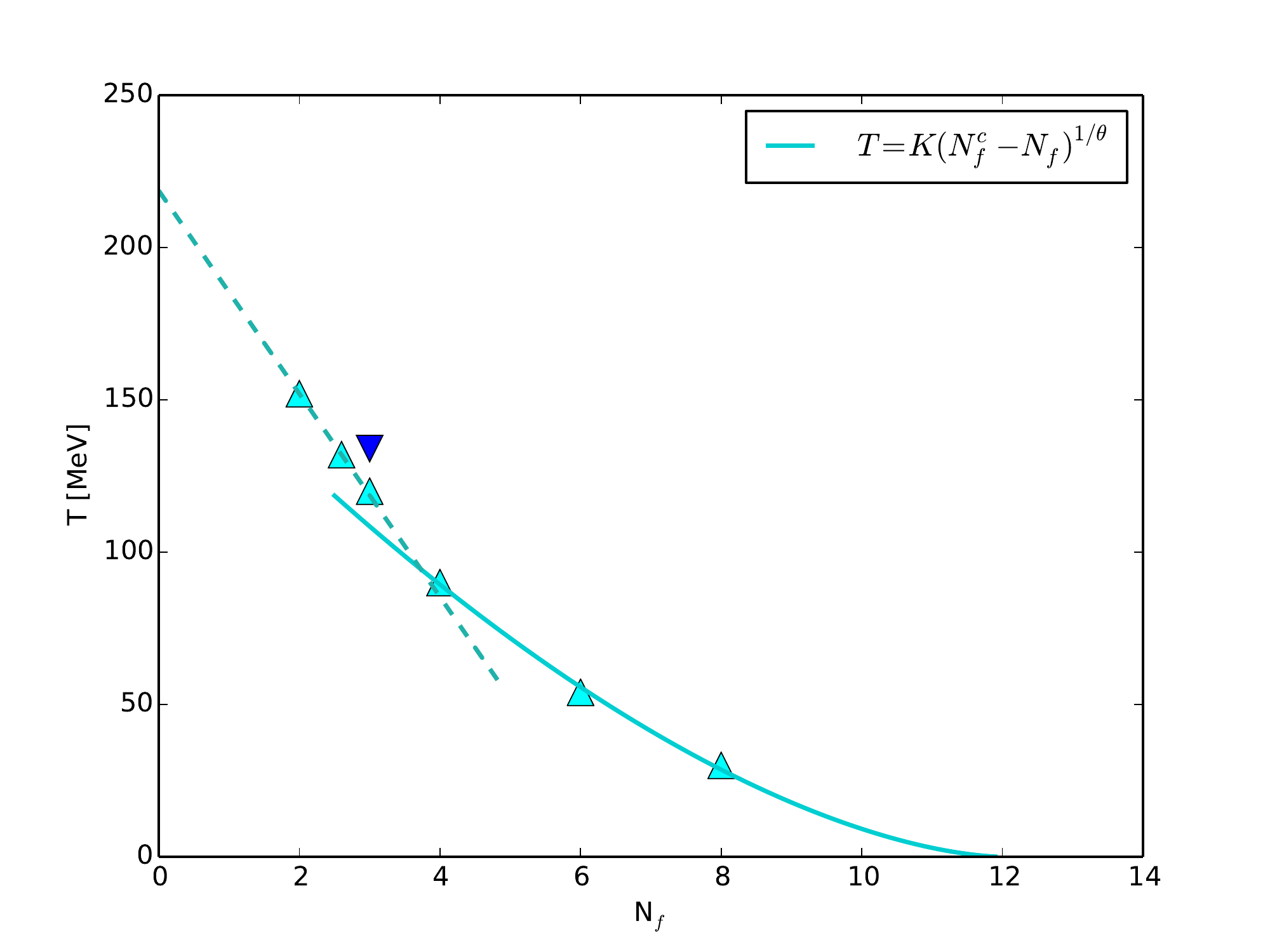}
    \caption{A sketchy view of the numerical results for the critical temperature~$T_c$ in the temperature, number of flavor plane. The theories with $N_f=2$ and $N_f = 2+1$ (marked as non-integer number of flavors) have been summarized in Sections~\ref{sec:sec3} and~\ref{sec:nf2p1}, $N_f = 3$ and $N_f = 4$ in Section~\ref{sec:nf3-4}, and larger $N_f$ in Section~\ref{sec:largenf}.} The approach to the conformal window for $N_f \simeq 12$ is apparent for $N_f \ge 4$. See text for details. 
    \label{fig:tcnf}
\end{figure}

We plot the result for the critical temperature in the chiral limit in the
$m_\pi, m_s, T$ space in Figure~\ref{fig:nf2nf3sum}, and in the
$N_f,T$ plane in Figure~\ref{fig:tcnf}. In the latter case, we
have used the input from Ref.~\cite{Braun:2010qs},
which predicts a linear behaviour of the critical
line for small $N_f$,
and an estimate of the critical temperature for $N_f=3$ in the chiral limit to convert the 
result in the chiral limit for light quarks, and a physical strange mass, to a non-integer number of flavor $N_f \approx 2.6$. 

\begin{figure*}[t]
    \centering
    \vskip -3 cm
    \includegraphics[width=12cm]{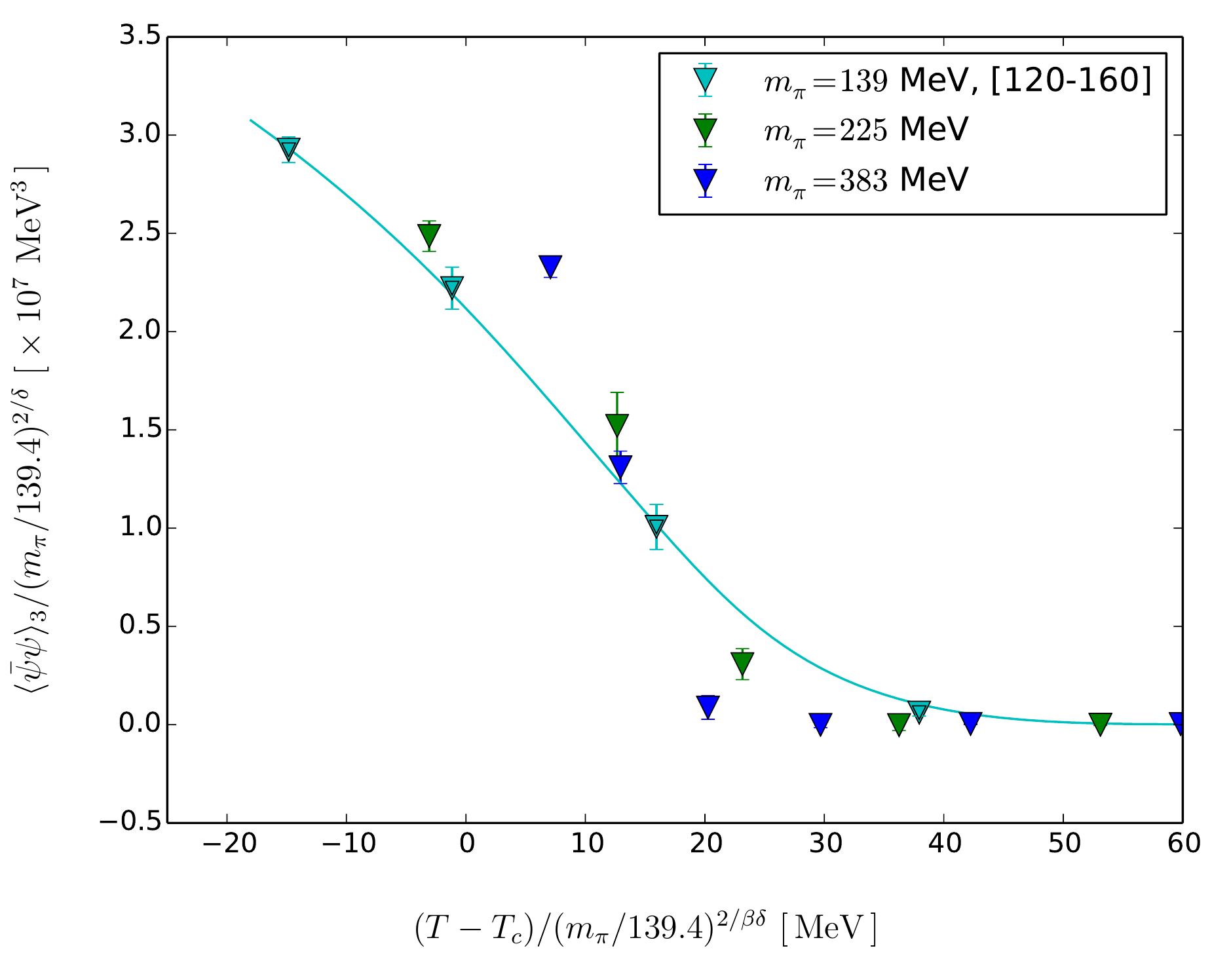}
    \caption{Emprical 3D $O(4)$ scaling with fixed $T_c = 138$ MeV; there is an apparent scale invariance, however the  universal EoS fitted for the physical pion mass ~-- computed with a fit in the interval [120--160 MeV] and marked as a continuous line -- does not fare well
    on the results for the other masses.}
    \label{fig:empirical}
\end{figure*}

Since  $\langle\bar\psi\psi\rangle_3$ 
 is free from additive renormalization, and the multiplicative renormalization is available, we can convert it to physical
units. This also allows us to attempt a semi-quantitative check of 
critical scaling.
One first simple way of doing this is to identify 
the scaling of the condensate at $T_c$: 
\begin{equation}
\langle \bar\psi\psi \rangle_3(m) \propto m_\pi^{2/\delta}.
\end{equation}
The results for the chiral condensate rescaled
by  $m_\pi^{2/\delta}$ should cross at the critical point in the chiral limit.  The curves for two lightest masses 
cross around $T = 138$ ~MeV~\cite{Kotov:2021rah}, which may be taken as a tentative
estimate of the critical temperature. 
We can then try to draw the (would be) scale invariant plot
$\langle \bar \psi \psi \rangle_3/{m_\pi^{2/\delta}}$ 
versus $(T - 138~\text{MeV})/{m_\pi^{\frac{2}{\beta\delta}}}$ for different masses. 
Indeed the results fall more or less on the same curve,
see Figure \ref{fig:empirical},
and we have observed that this approximate 
scaling behaviour degrades 
rapidly when $T_c$ is varied  by more than a couple of MeV around $T_c = 138$~MeV.
However, a fit to the 3D $O(4)$ Equation of State and a constrained $T_c = 138$ MeV works nicely only for the physical pion, see
the continuous line in Figure~\ref{fig:empirical}. This behaviour is reminiscent of that observed 
in Ref.~\cite{Braun_2011}, where an apparently good scaling is observed
at larger masses, which is, however, distinct from the predicted
three dimensional $O(4)$ scaling.  
In conclusion, 
after constraining  the critical temperature to the best
estimate in the chiral limit coming from the empirical universal scaling, we observe a qualitative scaling for the reduced variables,
but the would-be universal curve is clearly different from that
predicted by the 3D $O(4)$ universality. 

Next, we
fit to the 3D $O(4)$ Equation of State with an open critical temperature, and 
(pion mass dependent) scaling parameters. The fits are satisfactory,
but the would be critical temperature $T_c$ depends heavily on the pion mass: we find $T_c = 142(2), 159(3), 174(2)$ MeV, 
from light to heavy masses. Interestingly, for the physical pion
mass the result for the critical temperature in the chiral limit
is consistent with the estimate from the mass scaling  of the condensate. 

Summarizing:  we obtain a
good  scaling with a common temperature
$T_c = 138$~MeV, but at the price of violating the universal EoS. Or, 
we fit all the masses to the universal EoS, but at the price of forfeiting the parameters' scaling.
The only  consistency is for the  lowest pion mass, which may be
taken as an indication of the onset of the scaling behaviour for masses around the physical values. 

\begin{figure*}
\hskip -0.9cm    \includegraphics[width=8cm]{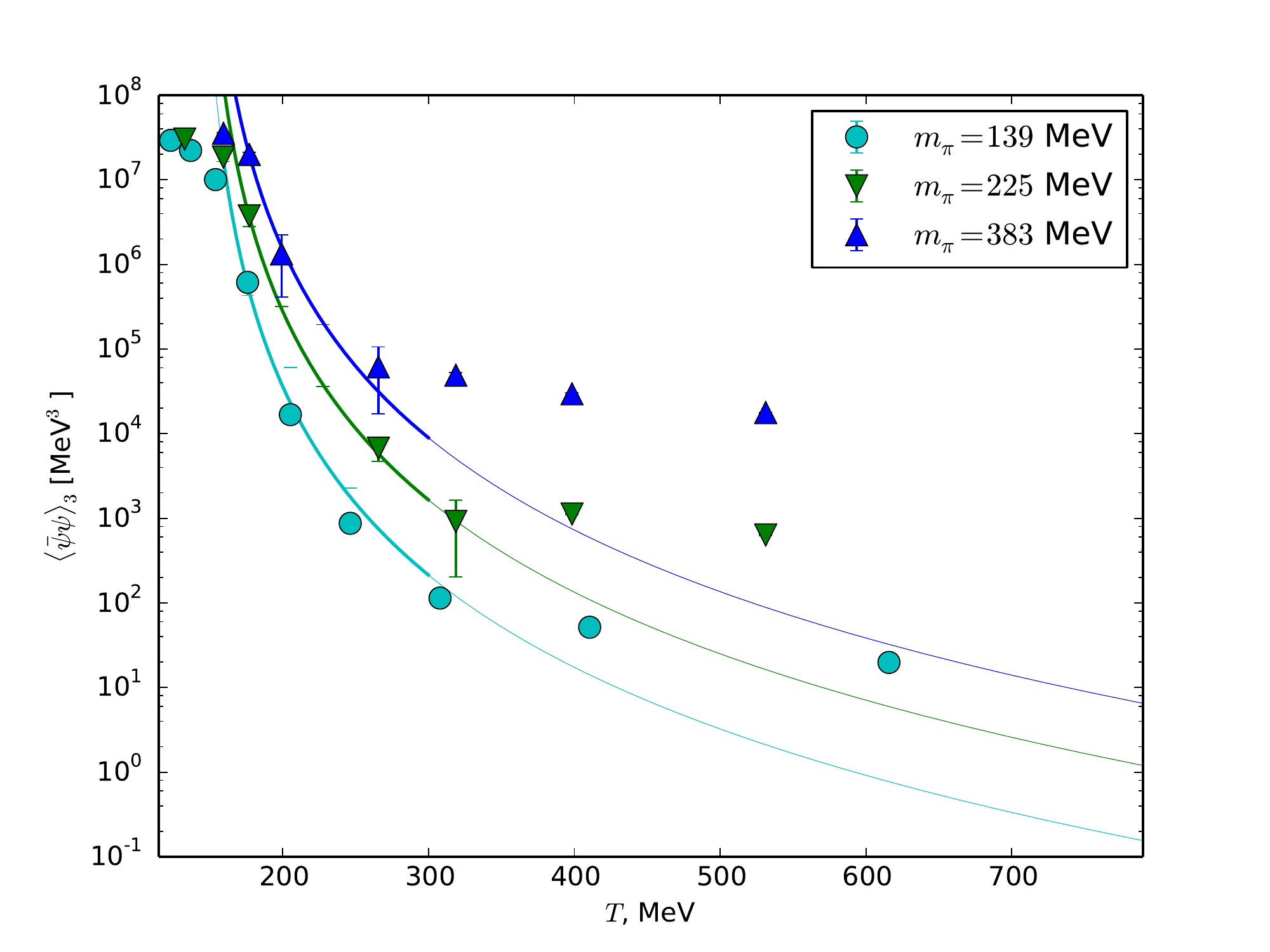}\hskip -0.8cm
    \includegraphics[width=8cm]{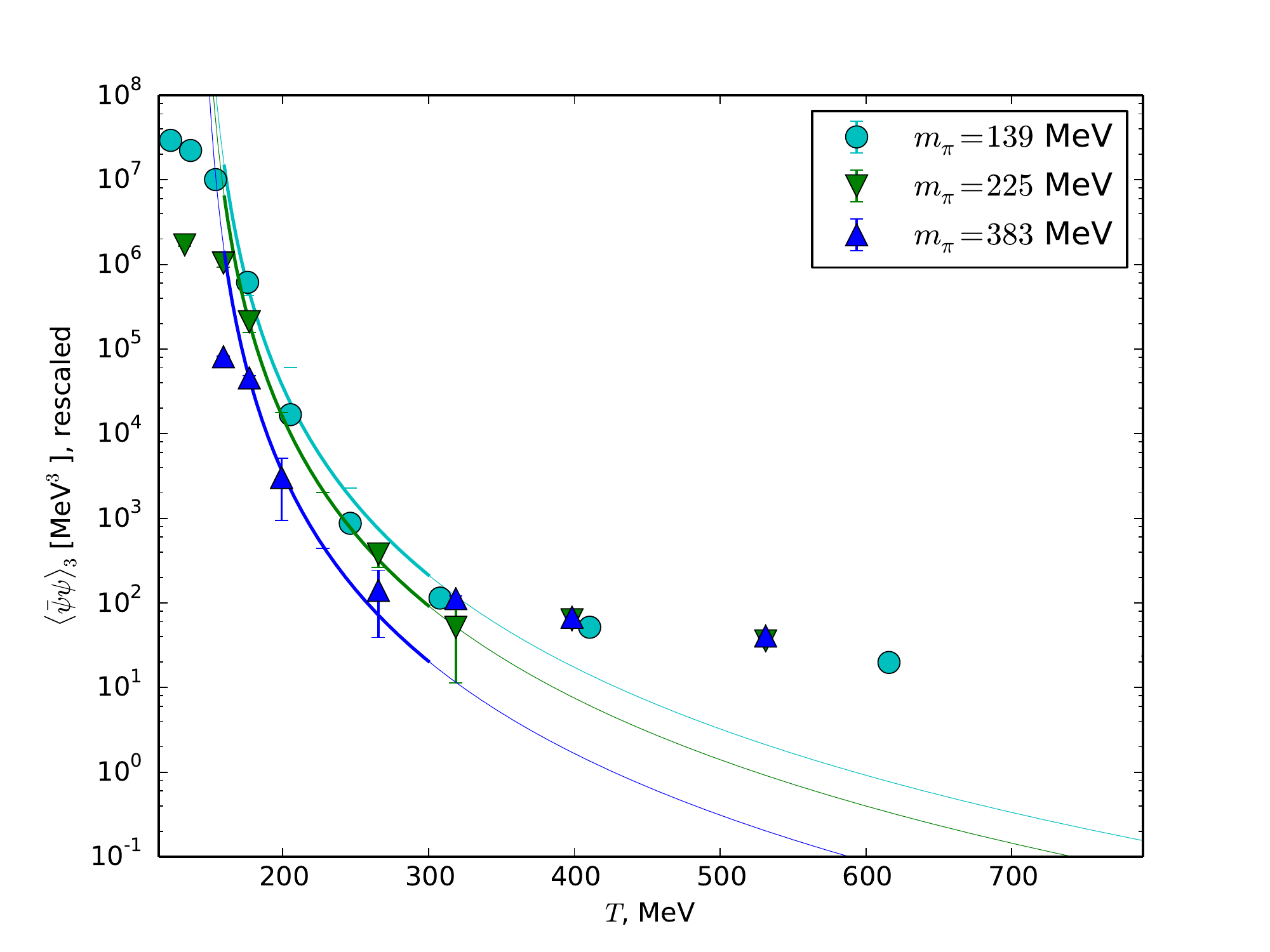}
    \caption{Fits to a constrained $O(4)$ behaviour: the results
in the interval of temperatures [160:300] MeV (marked bold) fare nicely through the data. For $T > 300$ MeV the behaviour is distinctly different. In the righthand plot (from Ref. \cite{Kotov:2021rah}) we show the data scaled according to $m^3 \simeq m_\pi^6$, the anticipated high temperature leading behaviour.}
    \label{fig:hight}
\end{figure*}

Finally, we consider the high temperature limit: in Figure \ref{fig:hight}, left, show fits to a constrained $O(4)$ behaviour, for our
preferred critical temperature in the chiral limit $T_c= 138$~MeV (the sensitivity to $T_c$ is very mild in this case): the results
in the interval of temperatures [160:300] MeV (marked bold) fare nicely through the data. For $T > 300$~MeV the behaviour is distinctly different:  in the right-hand plot (from Ref. \cite{Kotov:2021rah}) we show the data rescaled according to $m_q^3 \simeq m_\pi^6$, the anticipated high temperature leading behaviour, and indeed we see that the scaling is nicely satisfied 
above 300~MeV. This suggests that the temperature extent of the scaling window above $T_c$ extends up to about 300~MeV,
and then  a simple regular behaviour follows, unrelated with criticality. 
In a previous study~\cite{Burger:2018fvb,Lombardo:2020bvn} we have found that this is also the threshold for a behaviour
consistent with the Dilute Instanton Gas Approximation.

One final comment concerns  the $U(1)_A$ symmetry: given its prominent role, 
it is natural to resort to its analysis to try to shed more light on the symmetry pattern. But, again,
the problem remains open: 
 the current understanding is that 
it seems to be effectively restored above $T_c$ ~\cite{Ding:2020xlj,Kaczmarek:2021ser,Kaczmarek:2020sif, Aoki:2021qws,Aoki:2020noz,Mazur:2018pjw,Buchoff:2013nra,Suzuki:2020rla,Kanazawa:2015xna,Aoki:2012yj,Tomiya:2016jwr,Brandt:2019ksy,Brandt:2016daq,Cossu:2013uua,Chiu:2013wwa}, but there is no consensus on the restoration temperature. For instance, Ref.~\cite{Ding:2020xlj} finds the axial symmetry still broken at $T \simeq 1.6 T_c$, while Ref.~\cite{Brandt:2016daq} suggests a near-coincidence of axial and chiral transition. An interesting probe of the interrelation of the axial and chiral symmetry is the $\eta'$ meson, which seems to be well correlated with the chiral condensate also around $T_c$, favoring to some extent a close interrelation of the different symmetries \cite{Kotov:2019dby}.  

As a summary of this discussion, we plot the results in the
$m_\pi, m_s, T$ space in Figure~\ref{fig:nf2nf3sum}.

\section{\texorpdfstring{$N_f=3,4$}{nf3}: between the physical region and the pre-conformal window}
\label{sec:nf3-4}

Much of the effort in these cases focuses on the search for the
critical endpoint of the expected first order transition. 
Nice overviews of recent results can be found in \cite{Philipsen:2019rjq,deForcrand:2017cgb},
including an extensive bibliography. The 
 main conclusion (shared by all authors) is that the precise location of the
critical endpoint is hard to pinpoint, and very sensitive to the lattice discretization. Recent results from Ref.~\cite{Kuramashi:2020meg} indicate 
$m_\pi^c  \simeq  110$~MeV and $T_c \simeq 134(3)$~MeV. This value, rather close to the estimated critical temperature of the 
$N_f = 2+1$ flavor, is obviously an upper bound to the critical temperature in the chiral limit for the $N_f=3$ theory. Assuming - rather arbitrarily - 
that the slope of the critical first order line
is not too different from the slope of the pseudo-critical line of the $N_f = 2+1+1$ theory, one
may estimate a critical temperature for the $N_f=3$
theory at $T_c(N_f=3) \simeq  120$ MeV.
We note that some recent unpublished studies presented at the latest Lattice conference  indicate a lower value $T_c(N_f=3)\approx100$ MeV~\cite{Sharma:Lattice2021}.

The candidate endpoint, as well as the guess at the critical temperature in the chiral limit
are both marked in Figure \ref{fig:tcnf} as a blue and cyan triangles, respectively. 

Since most studies for $N_f=3$ have been carried out
with staggered fermions, a suggestion was made
\cite{deForcrand:2017cgb} that the rooting needed at $N_f=3$ may be the source of the strong lattice artifacts observed. This motivated an analysis of the $N_f=4$ theory, which is free from the rooting issue. However, also in this case it was not possible to locate the critical point with confidence.

In the most recent study~\cite{Cuteri:2021ikv} an extensive investigation  with unimproved staggered fermions
covering the whole range of $N_f=2$ to $N_f=8$ was reported. The results  suggest
that for all studied values of $N_f$ the first order region significantly shrinks upon taking the continuum limit and eventually the chiral transition in the chiral limit might be second-order (although a tiny first-order region cannot be excluded).

\section{\texorpdfstring{Large $N_f$}{largenf} }
\label{sec:largenf}

From now on, we approach the conformal window: a region of the
phase diagram where chiral symmetry remains unbroken also at zero 
temperature. Let us then 
take one step backwards, and ask: what triggers the breaking of the
$SU(N_f) \times SU(N_f)$ symmetry? In the following we briefly summarize the original model  calculations leading to the discovery of the conformal window \cite{Miransky:1996pd, Appelquist:1998rb}. It is clear that, since these phenomena are strongly-coupled, non-perturbative ones, ab-initio studies such as lattice QCD simulations are needed to confirm, or disprove,  analytic predictions.

Let us  consider
the renormalization group  equation for the running coupling:
\begin{equation}
\mu{\frac{\partial}{\partial \mu}} \alpha(\mu) = \beta(\alpha)
\equiv -b \alpha^2(\mu) -c\alpha^3(\mu) ...~,
\end{equation}
where $\alpha(\mu) = g^2(\mu)/4 \pi$. 
With $N$ colors and $N_f$ fermions in the fundamental representation 
\begin{eqnarray}
b &=& {\frac{1}{6 \pi}} \left( 11 N - 2 N_f\right), \\
c &=& {\frac{1}{24  \pi^2}} \left(34 N^2 - 10  N N_f - 3{\frac{N^2-1}{N}} N_f\right)~.
\end{eqnarray}

Hence,  the theory is asymptotically free if $b > 0$, i.e. 
$N_f < {\frac{11}{2}}N$, and it has
an infrared stable, non-trivial fixed point (FP)
$\alpha_* = -b/c $
if $b > 0$ and $c < 0$. This happens 
for    $\frac {34N^3}{ 13N^2 + 3}  < N_f < {\frac{11}{2}}N$, in short
$ N_f^\star < N_f < N_f^{\star \star}$.

With the infrared FP for  $ N_f^\star < N_f < N_f^{\star \star}$ 
the RG equation for the running coupling can be written as
\begin{equation}
b\log\left(\frac{q}{\mu}\right) = {\frac{1}{\alpha}} -
{\frac{1}{\alpha(\mu)}}- {\frac{1}{\alpha_*}}\log\left({\frac{\alpha
\left(\alpha(\mu) -
\alpha_*\right)}{\alpha(\mu)\left(\alpha-\alpha_*\right)}}\right)~,
\end{equation}
where $\alpha = \alpha(q)$.  

For $\alpha$, $\alpha(\mu)<\alpha_*$ we can introduce a scale defined by
\begin{equation}
\Lambda = \mu \exp\left[{{-1}\over{b \,\alpha_*}}
\log\left({{\alpha_*-\alpha(\mu)}
\over{\alpha(\mu)}}\right)-{{1}\over{b \alpha(\mu)}}\right]~,
\end{equation}
so that
$
{{1}\over{\alpha}} = b \log\left({{q}\over{\Lambda}}\right) +
{{1}\over{\alpha_*}}
\log\left({{\alpha}\over{\alpha_*-\alpha}}\right).
$
 Then,  for $q \gg \Lambda$ the running coupling displays the usual
perturbative behavior:
$
\alpha \approx {{1}\over{b \log\left({{q}\over{\Lambda}}\right)}}~,
\label{highalpha}
$ while  for $q \ll \Lambda$ it approaches the fixed point $\alpha_*$:
$
\alpha \approx {{\alpha_*}\over{1+ {{1}\over{e}}
\left({{q}\over{\Lambda}}\right)^{b \alpha_*}}}
\label{lowalpha}~.
$

These considerations, already present in the famous Banks-Zaks paper~\cite{Banks:1981nn}, lead
to the discovery of the conformal window~\cite{Appelquist:1998rb}, once one takes into
account the condition for chiral breaking. 
The analysis of  two-loop effective potential 
finds that chiral symmetry  breaking is favoured 
when 
\begin{equation}
\alpha_c \equiv {\frac{ \pi }{3 \, C_2(R)}}= 2 \pi \frac{N}{3\left(N^2-1\right)}~,
\end{equation}
where $C_2(R)$ is the quadratic Casimir of the representation.

Till there are no zeros of the beta function, this large value is always reached:  as long as $N_f$ is below the value  $N_f^c$ at
which $\alpha_{*} = \alpha_c$,  chiral symmetry is
spontaneously broken.  When the breaking happens, it washes out the IR fixed point and there is the usual running. 
For $N_f > N_f^c$ the chirally symmetric theory is infrared conformal \cite{Miransky:1996pd}, with anomalous dimension. The transition at $N_f^c$ is similar to the BKT one.
Below, but not too far from $N_f^c$, there is scale separation: in ordinary massless QCD dimensional transmutation generates a dimensionful parameters $\Lambda_{QCD}$
which is the natural mass scale of the theory. Close to the conformal window the coupling 'walks' rather than running, between two scales - above the UV scale there is the usual running, below the IR scale confinement sets in. In between the behaviour is near-conformal. This behaviour, known as scale separation (referring the the distinction between IR and UV scale)  offers~\cite{Lombardo:2014mda}  the possibility to build models for a composite Higgs. 
Lattice studies
have scrutinized in detail the model with $N_f=8$~\cite{Appelquist:2020xua,Witzel:2019jbe,Deuzeman:2008sc,DeGrand:2015zxa,Aoki:2013xza}, finding  evidences of
scale separation:  the lightest massive state, the scalar of the model, is suited for
phenomenology~-- it could be the Higgs meson.
We emphasize that at  $T=0$, it is very hard to distinguish a chirally broken
theory from a mass-deformed conformal theory,
see, for instance, Refs.~\cite{Golterman_2018,fodor2019tantalizing,Golterman_2020}.

Other vector states lie much above - this is where scale separation is needed -  which is why they haven't been observed so far~\cite{Cacciapaglia:2020kgq,DeGrand:2015zxa}.

Coming back to the main motivation of this writeup, and so to Figure~\ref{fig:tnf}, we are now interested in the thermal transition in the near-conformal region. The first complete sketch of Figure~\ref{fig:tnf} was obtained with FRG methods in Ref.~\cite{Braun:2006jd}. Lattice studies have focused on the very existence of the transition: indeed, not knowing exactly where the conformal phase begins, the observation of a thermal transition is {\em per se} an evidence of
a broken phase~\cite{Deuzeman:2008sc}, while within  the conformal window temperature merely breaks conformality,
and there is no thermal phase transition~\cite{Ishikawa:2013tua}.

A systematic study of the thermal phase transition as a function of the number of flavors has been carried out in Refs.  \cite{Miura:2011mc,Miura:2012zqa}. 
The pseudo-critical temperature has been identified by performing lattice simulations for $N_f=4,6,8$. After a suitable choice of a common scale among the different theories, it was possible to extrapolate $T_c(N_f)$  
to zero, thus identifying the candidate critical number of flavor. Here an
interesting issue appears: shall $T_c$ follow an essential scaling, as
expected of the conformal nature of the transition, or, rather, a power law scaling \cite{Braun:2010qs}? Again, the quality of the numerical results does not give a clear answer on the nature of the critical behaviour. However, again, luckily, the estimated critical number of flavor does not
depend on the parametrization chosen, within the largish errors~\cite{Lombardo:2014mda}.

In Figure~\ref{fig:tcnf} we show the results in the $N_f,T$ plane.  We
have used the input from Ref.~\cite{Braun:2010qs},
which predicts a linear behaviour of the critical
line for small $N_f$,
and an estimate of the critical temperature for $N_f=3$ in the chiral limit to convert the 
result in the chiral limit for light quarks, and a physical strange mass, to a non-integer number of flavor $N_f \approx 2.6$. 
The results for $N_f = 4,6,8 $ are normalized in such a way that $T_c(N_f=4)$
follows the linear behaviour predicted for a small number of flavors. The continuous line is the
predicted scaling of the critical temperature~\cite{Braun:2006jd}:
\begin{equation}
    T_c =K(N_f^c-N_f)^{-2b^2_0(N_f^c)/b_1(N_f^c)}
\end{equation}
with a fixed $N_f^c = 12$ (of course this does not depend on the normalization chosen). The exponent 
$-2b^2_0(N_f^c)/b_1(N_f^c) \simeq -1.64$ should
be contrasted  with the theoretical prediction 
$-2b^2_0(12)/b_1(12) = -1.05$ and would correspond
to $N_f^c \simeq 12.9$ \cite{Braun:2006jd}.

We are not aware of any theoretical modeling which explains how the first order behaviour for smaller $N_f$ eventually develops into the conformal transition. One possible scenario is that the second order $Z_2$ line, which terminates the first order region above the thermal line, shrinks to
zero at $N_f^c$. Another  possibility is a first order transition \cite{Antipin:2012sm,Sannino:2012wy}: in such a case the would-be critical number of flavor would correspond to a spinodal point, and the critical line would terminate at $8 < N_f^{1st} < 12$, where the lower bound stems
from the clean observation of chiral breaking in the eight flavor theory. 
One interesting information  emerging from the data is the strength  of the phase transition: it has been found that  it becomes stronger and stronger when approaching the conformal window~\cite{Miura:2012zqa,Shuryak:2013bxa,Philipsen:2019rjq}. 
Moreover,
at the critical point the coupling at the thermal transitions equals
the coupling at the infrared fixed point appearing there~\cite{Miura:2012zqa}. 
While the critical behaviour remains unclear, the dynamical scenario
seems thus well understood. In particular, 
the $N_f=8$ theory remains an interesting candidate for  physics beyond the Standard Model~\cite{Appelquist:2020xua}, and  its strong  first order transition may then be used to model a strong Electroweak transition and the generation of gravitational waves \cite{Miura:2018dsy}.

\section{Summary}

The study of the critical line of strong interactions has several
interesting points and remaining unknowns.

We started from Figure~\ref{fig:tnf} and we  progressively filled in the qualitative summary plot  Figure~\ref{fig:tcnf} with numerical results. The linear, low $N_f$   part of the critical line has been imposed, by aligning the $N_f = 2+1$ results with the $N_f =2$ and $N_f=3$,
and by suitably renormalizing  the results for large $N_f$. 

A detailed view for a small number of flavors is
given in Figure~\ref{fig:nf2nf3sum}. In that plot
 we have concentrated on the beginning of the chiral critical line, between
$N_f=2$ and $N_f=3$. We have reviewed our results for $N_f=2$ and for
$N_f=2+1+1$, with the strange flavor serving as an interpolator
between $N_f = 2$ and $N_f=3$.  We have discussed
the results at the physical point, as well as
the different scenarios for the chiral limit in the light sector  for $N_f=2$, and $N_f = 2+1$. We have identified a candidate scaling window for the 3D $O(4)$ theory: the physical 
pion mass maybe right at the onset of scaling, which extends up to temperatures of about $300$ MeV.

$N_f=3$ is an interesting unphysical model which would greatly help understanding the critical behaviour for $N_f=2+1$: we have briefly reviewed the status of the search of the endpoint for three
quarks of equal masses. Such endpoint would belong to the same $Z_2$ critical line as the $m_l=0, m_s^c$ point in Figure~\ref{fig:2ndtor}. Establishing (or ruling out) such a line would greatly contribute to building a consistent scenario for universality  in the physical case. 

We have then explored the large $N_f$ region, and discussed the approach to the conformal window. 
Clearly the results for the thermodynamics of these large number of flavors are much less developed 
than in the other cases, however there is
at least a good compatibility between the anticipated critical behaviour and the data, as well as between the estimated critical number of flavors
for the onset of conformality,  and the one inferred from the $T=0$ studies. It is confirmed that $N_f = 12$ is a subtle, borderline case, which justifies the use of  $N_f=8$ as a model for a walking theory, and related phenomenology. 

It remains to be understood how the transition changes its nature for first to second order, towards $N_f=2$. And, from the first order to BKT transition, at the onset of the conformal window, if indeed the BKT transition is realised~-- the possibility of a first order conformal transition  has been discussed as well~\cite{Antipin:2012sm,Sannino:2012wy}, as well as of a second order transition persisting for large $N_f$~\cite{Cuteri:2021ikv}, and this remains an open issue. 
In either cases  this transition may well happen for non-integer number of flavors,
or, correspondingly,  for a finite value of the interpolating mass
in the $N_f + 1$ model. The fate of the anomaly
plays an important  role in this discussion,
and a close comparison between numerical and analytic results
may well hold the key to a complete understanding of the properties
of the chiral line of strong interactions. 

\section*{Acknowledgements}
This work is partially supported by  STRONG-2020, a 
European Union’s Horizon 2020 research and innovation programme under grant agreement No. 824093.
The work of A.Yu.K. and A.T. was supported by RFBR grant 18-02-40126. A.T. acknowledges support from the "BASIS" foundation.
Numerical simulations have been carried out using computing resources of CINECA (based on the agreement between INFN and CINECA, on the ISCRA project IsB20), the supercomputer of Joint Institute for Nuclear Research ``Govorun'', and the computing resources of the federal collective usage center Complex for Simulation and Data Processing for Mega-science Facilities at NRC ``Kurchatov Institute'',~\url{http://ckp.nrcki.ru/}. 
\bibliographystyle{unsrt}
\bibliography{chiral,QCDphases}

\begin{thebibliography}{100}

\bibitem{Pisarski:1983ms}
Robert~D. Pisarski and Frank Wilczek.
\newblock {Remarks on the Chiral Phase Transition in Chromodynamics}.
\newblock {\em Phys. Rev. D}, 29:338--341, 1984.

\bibitem{Miransky:1996pd}
V.~A. Miransky and Koichi Yamawaki.
\newblock {Conformal phase transition in gauge theories}.
\newblock {\em Phys. Rev. D}, 55:5051--5066, 1997.
\newblock [Erratum: Phys.Rev.D 56, 3768 (1997)].

\bibitem{Rajagopal:1992qz}
Krishna Rajagopal and Frank Wilczek.
\newblock {Static and dynamic critical phenomena at a second order QCD phase
  transition}.
\newblock {\em Nucl. Phys. B}, 399:395--425, 1993.

\bibitem{Shuryak:2013bxa}
Edward Shuryak.
\newblock {QCD with many fermions and QCD topology}.
\newblock {\em J. Phys. Conf. Ser.}, 432:012022, 2013.

\bibitem{Philipsen:2019rjq}
Owe Philipsen.
\newblock {Constraining the phase diagram of QCD at finite temperature and
  density}.
\newblock {\em PoS}, LATTICE2019:273, 2019.

\bibitem{Cuteri:2018wci}
Francesca Cuteri, Owe Philipsen, and Alessandro Sciarra.
\newblock {Progress on the nature of the QCD thermal transition as a function
  of quark flavors and masses}.
\newblock {\em PoS}, LATTICE2018:170, 2018.

\bibitem{Cuteri:2017gci}
Francesca Cuteri, Owe Philipsen, and Alessandro Sciarra.
\newblock {QCD chiral phase transition from noninteger numbers of flavors}.
\newblock {\em Phys. Rev. D}, 97(11):114511, 2018.

\bibitem{MIRANSKY_2010}
V.~A. MIRANSKY.
\newblock Conformal phase transition in qcd like theories and beyond.
\newblock {\em International Journal of Modern Physics A},
  25(27n28):5105–5113, Nov 2010.

\bibitem{Braun:2010qs}
Jens Braun, Christian~S. Fischer, and Holger Gies.
\newblock {Beyond Miransky Scaling}.
\newblock {\em Phys. Rev. D}, 84:034045, 2011.

\bibitem{Antipin:2012sm}
Oleg Antipin, Matin Mojaza, and Francesco Sannino.
\newblock {Jumping out of the light-Higgs conformal window}.
\newblock {\em Phys. Rev. D}, 87(9):096005, 2013.

\bibitem{Sannino:2012wy}
Francesco Sannino.
\newblock {Jumping Dynamics}.
\newblock {\em Mod. Phys. Lett. A}, 28:1350127, 2013.

\bibitem{Ratti:2018ksb}
Claudia Ratti.
\newblock {Lattice QCD and heavy ion collisions: a review of recent progress}.
\newblock {\em Rept. Prog. Phys.}, 81(8):084301, 2018.

\bibitem{Ding:2020rtq}
Heng-Tong Ding.
\newblock {New developments in lattice QCD on equilibrium physics and phase
  diagram}.
\newblock In {\em {28th International Conference on Ultrarelativistic
  Nucleus-Nucleus Collisions}}, 2 2020.

\bibitem{Shuryak:1993ee}
Edward~V. Shuryak.
\newblock {Which chiral symmetry is restored in hot QCD?}
\newblock {\em Comments Nucl. Part. Phys.}, 21(4):235--248, 1994.

\bibitem{Pelissetto:2013hqa}
Andrea Pelissetto and Ettore Vicari.
\newblock {Relevance of the axial anomaly at the finite-temperature chiral
  transition in QCD}.
\newblock {\em Phys. Rev. D}, 88(10):105018, 2013.

\bibitem{Nicola:2020wxy}
Angel~G\'omez Nicola, Jacobo Ruiz~de Elvira, Andrea Vioque-Rodr\'\i{}guez, and
  David \'Alvarez-Herrero.
\newblock {The role of strangeness in chiral and $U(1)_A$ restoration}.
\newblock 12 2020.

\bibitem{Ding:2020xlj}
H.~T. Ding, S.~T. Li, Swagato Mukherjee, A.~Tomiya, X.~D. Wang, and Y.~Zhang.
\newblock {Correlated Dirac Eigenvalues and Axial Anomaly in Chiral Symmetric
  QCD}.
\newblock {\em Phys. Rev. Lett.}, 126(8):082001, 2021.

\bibitem{Kaczmarek:2021ser}
Olaf Kaczmarek, Lukas Mazur, and Sayantan Sharma.
\newblock {Eigenvalue spectra of QCD and the fate of $U_A(1)$ breaking towards
  the chiral limit}.
\newblock 2 2021.

\bibitem{Kaczmarek:2020sif}
Olaf Kaczmarek, Frithjof Karsch, Anirban Lahiri, Lukas Mazur, and Christian
  Schmidt.
\newblock {QCD phase transition in the chiral limit}.
\newblock 3 2020.

\bibitem{Aoki:2021qws}
S.~Aoki, Y.~Aoki, H.~Fukaya, S.~Hashimoto, C.~Rohrhofer, and K.~Suzuki.
\newblock {Role of axial U(1) anomaly in chiral susceptibility of QCD at high
  temperature}.
\newblock 3 2021.

\bibitem{Aoki:2020noz}
S.~Aoki, Y.~Aoki, G.~Cossu, H.~Fukaya, S.~Hashimoto, T.~Kaneko, C.~Rohrhofer,
  and K.~Suzuki.
\newblock {Study of the axial $U(1)$ anomaly at high temperature with lattice
  chiral fermions}.
\newblock {\em Phys. Rev. D}, 103(7):074506, 2021.

\bibitem{Mazur:2018pjw}
Lukas Mazur, Olaf Kaczmarek, Edwin Laermann, and Sayantan Sharma.
\newblock {The fate of axial U(1) in 2+1 flavor QCD towards the chiral limit}.
\newblock {\em PoS}, LATTICE2018:153, 2019.

\bibitem{Buchoff:2013nra}
Michael~I. Buchoff, Michael Cheng, Norman~H. Christ, H.-T. Ding, Chulwoo Jung,
  F.~Karsch, Zhongjie Lin, R.~D. Mawhinney, Swagato Mukherjee, P.~Petreczky,
  Dwight Renfrew, Chris Schroeder, P.~M. Vranas, and Hantao Yin.
\newblock Qcd chiral transition, $u(1{)}_{A}$ symmetry and the dirac spectrum
  using domain wall fermions.
\newblock {\em Phys. Rev. D}, 89:054514, Mar 2014.

\bibitem{Suzuki:2020rla}
Kei Suzuki, Sinya Aoki, Yasumichi Aoki, Guido Cossu, Hidenori Fukaya, Shoji
  Hashimoto, and Christian Rohrhofer.
\newblock {Axial U(1) symmetry and mesonic correlators at high temperature in
  $N_f=2$ lattice QCD}.
\newblock In {\em {37th International Symposium on Lattice Field Theory}}, 1
  2020.

\bibitem{Kanazawa:2015xna}
Takuya Kanazawa and Naoki Yamamoto.
\newblock {U (1) axial symmetry and Dirac spectra in QCD at high temperature}.
\newblock {\em JHEP}, 01:141, 2016.

\bibitem{Aoki:2012yj}
Sinya Aoki, Hidenori Fukaya, and Yusuke Taniguchi.
\newblock {Chiral symmetry restoration, eigenvalue density of Dirac operator
  and axial U(1) anomaly at finite temperature}.
\newblock {\em Phys. Rev. D}, 86:114512, 2012.

\bibitem{Tomiya:2016jwr}
A.~Tomiya, G.~Cossu, S.~Aoki, H.~Fukaya, S.~Hashimoto, T.~Kaneko, and J.~Noaki.
\newblock {Evidence of effective axial U(1) symmetry restoration at high
  temperature QCD}.
\newblock {\em Phys. Rev. D}, 96(3):034509, 2017.
\newblock [Addendum: Phys.Rev.D 96, 079902 (2017)].

\bibitem{Brandt:2019ksy}
Bastian~B. Brandt, Marco Cè, Anthony Francis, Tim Harris, Harvey~B. Meyer,
  Hartmut Wittig, and Owe Philipsen.
\newblock {Testing the strength of the $\text{U}_A(1)$ anomaly at the chiral
  phase transition in two-flavour QCD}.
\newblock {\em PoS}, CD2018:055, 2019.

\bibitem{Brandt:2016daq}
Bastian~B. Brandt, Anthony Francis, Harvey~B. Meyer, Owe Philipsen, Daniel
  Robaina, and Hartmut Wittig.
\newblock {On the strength of the $U_A(1)$ anomaly at the chiral phase
  transition in $N_f=2$ QCD}.
\newblock {\em JHEP}, 12:158, 2016.

\bibitem{Cossu:2013uua}
Guido Cossu, Sinya Aoki, Hidenori Fukaya, Shoji Hashimoto, Takashi Kaneko,
  Hideo Matsufuru, and Jun-Ichi Noaki.
\newblock {Finite temperature study of the axial U(1) symmetry on the lattice
  with overlap fermion formulation}.
\newblock {\em Phys. Rev. D}, 87(11):114514, 2013.
\newblock [Erratum: Phys.Rev.D 88, 019901 (2013)].

\bibitem{Chiu:2013wwa}
Ting-Wai Chiu, Wen-Ping Chen, Yu-Chih Chen, Han-Yi Chou, and Tung-Han Hsieh.
\newblock {Chiral symmetry and axial U(1) symmetry in finite temperature QCD
  with domain-wall fermion}.
\newblock {\em PoS}, LATTICE2013:165, 2014.

\bibitem{Kotov:2021rah}
A.~Yu. Kotov, M.~P. Lombardo, and A.~Trunin.
\newblock {QCD transition at the physical point, and its scaling window from
  twisted mass Wilson fermions}.
\newblock 5 2021.

\bibitem{Kotov:2020hzm}
Andrey~{\relax Yu}. Kotov, Maria~Paola Lombardo, and Anton~M. Trunin.
\newblock {Finite temperature QCD with $N_f=2+1+1$ Wilson twisted mass fermions
  at physical pion, strange and charm masses}.
\newblock {\em Eur. Phys. J.}, A56(8):203, 2020.

\bibitem{Kotov:2019dby}
Andrey~Yu. Kotov, Maria~Paola Lombardo, and Anton~M. Trunin.
\newblock {Fate of the $\eta^{'}$ in the quark gluon plasma}.
\newblock {\em Phys. Lett. B}, 794:83--88, 2019.

\bibitem{Lombardo:2014mda}
Maria~Paola Lombardo, Kohtaroh Miura, Tiago~J. Nunes~da Silva, and Elisabetta
  Pallante.
\newblock {One, two, zero: Scales of strong interactions}.
\newblock {\em Int. J. Mod. Phys. A}, 29(25):1445007, 2014.

\bibitem{Miura:2012zqa}
Kohtaroh Miura and Maria~Paola Lombardo.
\newblock {Lattice Monte-Carlo study of pre-conformal dynamics in strongly
  flavoured QCD in the light of the chiral phase transition at finite
  temperature}.
\newblock {\em Nucl. Phys. B}, 871:52--81, 2013.

\bibitem{Miura:2011mc}
Kohtaroh Miura, Maria~Paola Lombardo, and Elisabetta Pallante.
\newblock {Chiral phase transition at finite temperature and conformal dynamics
  in large Nf QCD}.
\newblock {\em Phys. Lett. B}, 710:676--682, 2012.

\bibitem{Cacciapaglia:2020kgq}
Giacomo Cacciapaglia, Claudio Pica, and Francesco Sannino.
\newblock {Fundamental Composite Dynamics: A Review}.
\newblock {\em Phys. Rept.}, 877:1--70, 2020.

\bibitem{Pelissetto:2000ek}
Andrea Pelissetto and Ettore Vicari.
\newblock {Critical phenomena and renormalization group theory}.
\newblock {\em Phys. Rept.}, 368:549--727, 2002.

\bibitem{Engels:1999wf}
Jurgen Engels and Tereza Mendes.
\newblock {Goldstone mode effects and scaling function for the
  three-dimensional O(4) model}.
\newblock {\em Nucl. Phys.}, B572:289--304, 2000.

\bibitem{Ginzburg}
V.L. Ginzburg.
\newblock {\em Fiz. Tverd. Tela}, 2:1824, 1960.

\bibitem{Fernandez:1992ns}
L.~A. Fernandez, J.~J. Ruiz-Lorenzo, M.~P. Lombardo, and A.~Tarancon.
\newblock {Weak first order transitions: The Two-dimensional Potts model}.
\newblock {\em Phys. Lett. B}, 277:485--490, 1992.

\bibitem{Hohenberg_2015}
P.C. Hohenberg and A.P. Krekhov.
\newblock An introduction to the ginzburg–landau theory of phase transitions
  and nonequilibrium patterns.
\newblock {\em Physics Reports}, 572:1–42, Apr 2015.

\bibitem{Engels:2011km}
J.~Engels and F.~Karsch.
\newblock {The scaling functions of the free energy density and its derivatives
  for the 3d O(4) model}.
\newblock {\em Phys. Rev.}, D85:094506, 2012.

\bibitem{Kocic:1992is}
Aleksandar Kocic, John~B. Kogut, and Maria-Paola Lombardo.
\newblock {Universal properties of chiral symmetry breaking}.
\newblock {\em Nucl. Phys.}, B398:376--404, 1993.

\bibitem{Karsch:1994hm}
Frithjof Karsch and Edwin Laermann.
\newblock {Susceptibilities, the specific heat and a cumulant in two flavor
  QCD}.
\newblock {\em Phys. Rev. D}, 50:6954--6962, 1994.

\bibitem{Ding:2019prx}
H.T. Ding et~al.
\newblock {Chiral Phase Transition Temperature in ( 2+1 )-Flavor QCD}.
\newblock {\em Phys. Rev. Lett.}, 123(6):062002, 2019.

\bibitem{Kogut:1998rh}
J.~B. Kogut, J.~F. Lagae, and D.~K. Sinclair.
\newblock {Topology, fermionic zero modes and flavor singlet correlators in
  finite temperature QCD}.
\newblock {\em Phys. Rev.}, D58:054504, 1998.

\bibitem{Unger:2010wcq}
Wolfgang Unger.
\newblock {\em {The chiral phase transition of QCD with 2+1 flavors: a lattice
  study on Goldstone modes and universal scaling}}.
\newblock PhD thesis, U. Bielefeld (main), 2010.

\bibitem{Kocic:1995nf}
Aleksandar Kocic and John~B. Kogut.
\newblock {Phase transitions at finite temperature and dimensional reduction
  for fermions and bosons}.
\newblock {\em Nucl. Phys. B}, 455:229--273, 1995.

\bibitem{Caselle:2020tjz}
Michele Caselle and Marianna Sorba.
\newblock {Charting the scaling region of the Ising universality class in two
  and three dimensions}.
\newblock {\em Phys. Rev. D}, 102(1):014505, 2020.

\bibitem{Pisarski}
{We thank Rob Pisarski for discussions on this point}.

\bibitem{Miura:2018dsy}
Kohtaroh Miura, Hiroshi Ohki, Saeko Otani, and Koichi Yamawaki.
\newblock {Gravitational Waves from Walking Technicolor}.
\newblock {\em JHEP}, 10:194, 2019.

\bibitem{PhysRevLett.108.255703}
Tomoaki Nogawa, Takehisa Hasegawa, and Koji Nemoto.
\newblock Generalized scaling theory for critical phenomena including essential
  singularities and infinite dimensionality.
\newblock {\em Phys. Rev. Lett.}, 108:255703, Jun 2012.

\bibitem{Vicari:2008jw}
Ettore Vicari and Haralambos Panagopoulos.
\newblock {Theta dependence of SU(N) gauge theories in the presence of a
  topological term}.
\newblock {\em Phys. Rept.}, 470:93--150, 2009.

\bibitem{Burger:2011zc}
Florian Burger, Ernst-Michael Ilgenfritz, Malik Kirchner, Maria~Paola Lombardo,
  Michael Müller-Preussker, Owe Philipsen, Carsten Urbach, and Lars
  Zeidlewicz.
\newblock {Thermal QCD transition with two flavors of twisted mass fermions}.
\newblock {\em Phys. Rev. D}, 87(7):074508, 2013.

\bibitem{Braun:2020ada}
Jens Braun, Wei-Jie Fu, Jan~M. Pawlowski, Fabian Rennecke, Daniel Rosenblüh,
  and Shi Yin.
\newblock {Chiral Susceptibility in (2+1)-flavour QCD}.
\newblock 3 2020.

\bibitem{Bazavov:2018mes}
A.~Bazavov et~al.
\newblock {Chiral crossover in QCD at zero and non-zero chemical potentials}.
\newblock {\em Phys. Lett.}, B795:15--21, 2019.

\bibitem{Aarts:2019hrg}
Gert Aarts et~al.
\newblock {Spectral quantities in thermal QCD: a progress report from the
  FASTSUM collaboration}.
\newblock In {\em {37th International Symposium on Lattice Field Theory}},
  2019.

\bibitem{Aarts:2020vyb}
G.~Aarts et~al.
\newblock {Properties of the QCD thermal transition with $N_f=2+1$ flavours of
  Wilson quark}.
\newblock 7 2020.

\bibitem{Borsanyi_2020}
Szabolcs Borsanyi, Zoltan Fodor, Jana~N. Guenther, Ruben Kara, Sandor~D. Katz,
  Paolo Parotto, Attila Pasztor, Claudia Ratti, and Kálman~K. Szabó.
\newblock Qcd crossover at finite chemical potential from lattice simulations.
\newblock {\em Physical Review Letters}, 125(5), Jul 2020.

\bibitem{Braun_2011}
Jens Braun, Bertram Klein, and Piotr Piasecki.
\newblock On the scaling behavior of the chiral phase transition in qcd
  in finite and infinite volume.
\newblock {\em The European Physical Journal C}, 71(3), Mar 2011.

\bibitem{Burger:2018fvb}
Florian Burger, Ernst-Michael Ilgenfritz, Maria~Paola Lombardo, and Anton
  Trunin.
\newblock {Chiral observables and topology in hot QCD with two families of
  quarks}.
\newblock {\em Phys. Rev. D}, 98(9):094501, 2018.

\bibitem{Lombardo:2020bvn}
Maria~Paola Lombardo and Anton Trunin.
\newblock {Topology and axions in QCD}.
\newblock {\em Int. J. Mod. Phys. A}, 35(20):2030010, 2020.

\bibitem{deForcrand:2017cgb}
Philippe de~Forcrand and Massimo D'Elia.
\newblock {Continuum limit and universality of the Columbia plot}.
\newblock {\em PoS}, LATTICE2016:081, 2017.

\bibitem{Kuramashi:2020meg}
Yoshinobu Kuramashi, Yoshifumi Nakamura, Hiroshi Ohno, and Shinji Takeda.
\newblock {Nature of the phase transition for finite temperature $N_{\rm f}=3$
  QCD with nonperturbatively O($a$) improved Wilson fermions at $N_{\rm
  t}=12$}.
\newblock {\em Phys. Rev. D}, 101(5):054509, 2020.

\bibitem{Sharma:Lattice2021}
S.~Sharma et~al.
\newblock Talk at lattice 2021, to appear in the proceedings, 2021.

\bibitem{Cuteri:2021ikv}
Francesca Cuteri, Owe Philipsen, and Alessandro Sciarra.
\newblock {On the order of the QCD chiral phase transition for different
  numbers of quark flavours}.
\newblock 7 2021.

\bibitem{Appelquist:1998rb}
Thomas Appelquist, Anuradha Ratnaweera, John Terning, and L.~C.~R.
  Wijewardhana.
\newblock {The Phase structure of an SU(N) gauge theory with N(f) flavors}.
\newblock {\em Phys. Rev. D}, 58:105017, 1998.

\bibitem{Banks:1981nn}
Tom Banks and A.~Zaks.
\newblock {On the Phase Structure of Vector-Like Gauge Theories with Massless
  Fermions}.
\newblock {\em Nucl. Phys. B}, 196:189--204, 1982.

\bibitem{Appelquist:2020xua}
Thomas Appelquist et~al.
\newblock {Near-conformal dynamics in a chirally broken system}.
\newblock {\em Phys. Rev. D}, 103(1):014504, 2021.

\bibitem{Witzel:2019jbe}
Oliver Witzel.
\newblock {Review on Composite Higgs Models}.
\newblock {\em PoS}, LATTICE2018:006, 2019.

\bibitem{Deuzeman:2008sc}
Albert Deuzeman, Maria~Paola Lombardo, and Elisabetta Pallante.
\newblock {The Physics of eight flavours}.
\newblock {\em Phys. Lett. B}, 670:41--48, 2008.

\bibitem{DeGrand:2015zxa}
Thomas DeGrand.
\newblock {Lattice tests of beyond Standard Model dynamics}.
\newblock {\em Rev. Mod. Phys.}, 88:015001, 2016.

\bibitem{Aoki:2013xza}
Yasumichi Aoki, Tatsumi Aoyama, Masafumi Kurachi, Toshihide Maskawa, Kei-ichi
  Nagai, Hiroshi Ohki, Akihiro Shibata, Koichi Yamawaki, and Takeshi Yamazaki.
\newblock {Walking signals in $N_f=8$ QCD on the lattice}.
\newblock {\em Phys. Rev. D}, 87(9):094511, 2013.

\bibitem{Golterman_2018}
Maarten Golterman and Yigal Shamir.
\newblock Large-mass regime of the dilaton-pion low-energy effective theory.
\newblock {\em Physical Review D}, 98(5), Sep 2018.

\bibitem{fodor2019tantalizing}
Zoltan Fodor, Kieran Holland, Julius Kuti, and Chik~Him Wong.
\newblock Tantalizing dilaton tests from a near-conformal eft, 2019.

\bibitem{Golterman_2020}
Maarten Golterman, Ethan~T. Neil, and Yigal Shamir.
\newblock Application of dilaton chiral perturbation theory to nf=8 , su(3)
  spectral data.
\newblock {\em Physical Review D}, 102(3), Aug 2020.

\bibitem{Braun:2006jd}
Jens Braun and Holger Gies.
\newblock {Chiral phase boundary of QCD at finite temperature}.
\newblock {\em JHEP}, 06:024, 2006.

\bibitem{Ishikawa:2013tua}
K.~I. Ishikawa, Y.~Iwasaki, Yu~Nakayama, and T.~Yoshie.
\newblock {Global Structure of Conformal Theories in the SU(3) Gauge Theory}.
\newblock {\em Phys. Rev. D}, 89(11):114503, 2014.

\end{thebibliography}
\end{document}